\documentclass[12pt,preprint,natbib]{aastex}

\def\gsimeq{{_>\atop^{\sim}}}
\def\lsimeq{{_<\atop^{\sim}}}

\lefthead{F. Pozzi, C. Gruppioni, S. Oliver et al.}
\righthead{The Mid-IR luminosity function of galaxies}

\begin{document}

\title{The Mid-IR luminosity function of galaxies \\ in the ELAIS Southern fields}

\author{F. Pozzi$^{1,2}$, C. Gruppioni$^{2}$, S. Oliver$^{3}$,
  I. Matute$^{4,5}$, F. La Franca$^{5}$, C. Lari$^{6}$, G. Zamorani$^{2}$, A. Franceschini$^{7}$, M. Rowan-Robinson$^{8}$}


\altaffiltext{1}{Dipartimento di Astronomia, Universit\`a di Bologna, viale Berti Pichat
6, I--40127 Bologna, Italy}

\altaffiltext{2}{INAF - Osservatorio Astronomico di Bologna, via
  Ranzani 1, I--40127 Bologna, Italy}
\altaffiltext{3}{Astronomy Centre, Department of Physics \& Astronomy, 
School of Science and Technology, University of Sussex, Brighton, BN1 9QJ, UK}
\altaffiltext{4}{Max-Planck-Institut f\"ur extraterrestrische Physik, Postfach
  1312, D-85741 Garching, Germany}
\altaffiltext{5}{Dipartimento di Fisica, Universit\`a 'Roma Tre', via della Vasca Navale
84, I-00146 Roma, Italy}
\altaffiltext{6}{Istituto di Radioastronomia del CNR, via Gobetti 101, I-40129 Bologna,
Italy }
\altaffiltext{7}{Dipartimento di Astronomia, Universit\`a di Padova, vicolo
dell'Osservatorio 2, I-35122 Padova, Italy}
\altaffiltext{8}{Astrophysics Group, Blackett Laboratory, Imperial College of
  Science, Technology and Medicine, Prince Consort Road, London SW7
  2BW, UK }


\begin{abstract}
We present the first determination of the 15$\mu$m luminosity function
of galaxies from the European Large Area ISO survey (ELAIS) southern fields. We have adopted a
new criterion to separate the quiescent, non-evolving and the starburst, evolving populations
based on the ratio of mid-infrared to optical luminosities.
Strong evolution is suggested by our data for the starburst galaxy population,
while normal spiral galaxies are consistent with no evolution. The starburst population must evolve both in
luminosity and in density with rates of the order $L(z) \propto (1+z)^{3.5}$ and
$\rho(z) \propto (1+z)^{3.8}$ up to $z \sim 1$. The evolutionary parameters of
our model have been tested by comparing the model predictions with other
observables, like source counts at all flux density levels (from 0.1 to 300
mJy) and redshift distributions and luminosity functions at high-$z$ (0.7 $< z <$ 1.0
from HDF-N data). The agreement between our model predictions and the observed data 
is remarkably good. We use our data to estimate the star-formation
  density of the Universe up to z=0.4 and we use the luminosity function model to
  predict the trend of the star-formation history up to $z=1$. 

\end{abstract}





%

\section{Introduction}
\label{intro_sec}
The extragalactic background light shows that the emission from
galaxies at infrared and sub-millimeter wavelengths is an energetically
significant component of the Universe. This emission originates from
star-formation activity and active galactic nuclei. The precise
contribution from each type of activity is still debated. It is thus
important to our understanding of galaxy and AGN formation to study
those populations that emit a substantial amount of light at infrared
(IR) wavelengths in their rest frame.
In particular, data from deep {\it Infrared Space Observatory} (ISO) 
surveys at 15 $\mu$m (i.e. \citealp{1999A&A...351L..37E}; \citealp{1999ApJ...517..148F};
\citealp{2001MNRAS.325.1173L}; \citealp{2003A&A...407..791M}) seem to require strong
evolution for galaxies emitting in the infrared wavebands. This result, supported
also by the detection of a substantial cosmic Infrared Background
in the 140 $\mu$m - 1 mm range \citep{1996A&A...308L...5P,1998ApJ...508...25H,1999A&A...344..322L}, 
has stimulated the development of several evolutionary models for IR galaxies
(i.e. \citealp{2001ApJ...549..745R}, \citealp{2001A&A...378....1F},
\citealp{2001ApJ...556..562C}, \citealp{2003ApJ...587...90X}).
 All these models fit with different degrees of success the IR/sub-millimeter source counts and the cosmic Infrared Background
within the present uncertainty limits, but suffer of parameter degeneracy and none of them 
is based on a local luminosity function (LLF) obtained from 15-$\mu$m
data, being all extrapolated from different IR wavelengths (12, 25, 60
$\mu$m).

So far, complete spectroscopic samples of 15-$\mu$m sources have been
obtained only in small fields (i.e. HDF-N: \citealp{1999usis.conf.1023A}; HDF-S:
\citealp{2002MNRAS.332..549M}, \citealp{2003A&A...403..501F}), too small and too
deep to allow a detailed study of the local luminosity function.

The ELAIS survey is the largest open time project conducted by ISO \citep{2000MNRAS.316..749O}, mapping an area of $\approx$ 12 deg$^2$ at 15
$\mu$m and 90 $\mu$m.  The final, band-merged ELAIS catalogue has
recently been completed \citep{MRR03}. The spectroscopic data is
most complete in the southern fields and in this paper we present an
analysis of the 15-$\mu$m luminosity function derived from these data.
This is the first determination of the 15-$\mu$m luminosity function 
and its evolution, constrained by all the available observables in this band
(source counts from IRAS to the deepest ISOCAM flux densities; redshift
distributions at low and high-$z$ using both data from ELAIS and data
from the deeper HDF-N survey). The model fitting the LF is then used to estimate
the star-formation history of the Universe.

This paper is structured as follows. In Section \ref{sample_sec} we present
our data sample. In Section \ref{kcorr_sec} we discuss the adopted IR and optical
K-corrections. In Section \ref{fdl_sec} we show the method used to
compute the 15-$\mu$m LF and present the results. In Section \ref{llf_section}
we compare our LLF determination with previous ones. In Section 
\ref{evolution_sec} we discuss the evolution rates derived from our data, 
and compare the model predictions with the observable constraints and with other
literature models. 
In Section \ref{concl_sec} we present our 
conclusions.

Throughout this paper we will assume $H_0 = 75$ km s$^{-1}$ Mpc$^{-1}$, 
$\Omega_m = 0.3$ and $\Omega_{\Lambda} = 0.7$, unless explicitly stated.

\section{The data samples}
\label{sample_sec}                           
\subsection{The parent catalogues}                     
Our analysis uses the southern ELAIS fields, which have more complete
spectroscopy than the northern fields at the present time. S1 and S2
are the main survey fields in the southern hemisphere: S1 is centered at
$\alpha(2000)= 00^{h}34^{m} 44.4 ^{s}$, $\delta(2000)=-43^{\circ} 28'
12''$ and covers an area of 2$^{\circ}{\times}2^{\circ}$, while S2 is
centered at $\alpha(2000) = 05^h 02^m 24.5^s, \delta(2000) =
-30^{\circ} 36^{\prime} 00^{\prime \prime}$ and covers an area of
$21^{\prime}{\times}21^{\prime}$.

The 15-$\mu$m data in these fields have been reduced using the {\it LARI technique} described in detail in \cite{2001MNRAS.325.1173L}
 and the 15-$\mu$m catalogues obtained in the two fields are given by
 \cite{2001MNRAS.325.1173L} and \cite{2003MNRAS.343.1348P}, respectively. The source counts
derived from the main field S1 are presented and discussed in
\cite{2002MNRAS.335..831G}. The whole S1 and S2 areas
have been surveyed in the radio down to $S_{1.4GHz}\approx{0.2}$ and
0.13 mJy respectively. The radio data analysis is presented in
\cite{1999MNRAS.305..297G} and Ciliegi et al. (in preparation).

S1 contains 462 sources detected at 15-$\mu$m, 406 of
which constitute a highly reliable sample which can be used in 
a conservative statistical analysis (see \citealp{fabio03}), while
S2 contains 43 15-$\mu$m sources.  Optical photometric follow-up has been
 obtained in S1 with the ESO/Danish 1.5m Telescope (to $R\sim$ 23) and in S2 with
the ESO WFI/2.2m Telescope in $U$, $B$ and $I$ (to $I\sim$22) and
SOFI/NTT in $K^{\prime}$ (to $K^{\prime}\sim$ 18.75). The
spectroscopic follow-up was carried out at the 2dF/AAT and ESO 3.6m,
NTT and Danish 1.5 Telescopes. In S1, $\sim$81 \% (328/406) of the
15-$\mu$m sources have an optical counterpart in the $R$
band (taking into account all the optical identification, sources to $R\sim$23
plus 6 sources with 23 $< R <$ 24.3, see \citealp{fabio03}), while in S2, where the identification has been
done in the $I$ band, the percentage is higher
($\sim$90 \%, 39/43). In S1, the spectroscopic redshift completeness is
$\sim$88 \% of identified objects and $\sim$71 \% of the whole sample (290 objects),
while in S2 the corresponding figures are $\sim$77 \% and
$\sim$70 \% (30 objects).

\subsection{The spectroscopic sample}
\label{spectra_sec}

To study the 15-$\mu$m luminosity function and its evolution we
restrict our analysis to 15-$\mu$m sources with optical counterparts
in the ranges of magnitude with higher spectroscopic identification
completeness. Recalling the
definition given in \cite{fabio03} of the S1$\_$rest region
(the shallower area) and of the S1$\_$5 region (the deeper
part), we consider the following objects: in S1$\_$rest all the sources down
to $R\le20.5$ (97 \%  complete, 216/223); in S1$\_$5 the sources down to
$R\le21.6$ (97 \%  complete, 73/75); in S2 the sources down to
$R\le22.6$ and with 15-$\mu$m fluxes $\ge$1 mJy (100 \%  complete,
18/18). In S2, the magnitude limit of $I \simeq 22$ can be converted
(for uniformity with S1) to a limit in $R$ magnitude, by assuming a mean
colour $R-I \sim 0.6$ (found for our galaxies). This leads to an $R$ limit of $\sim 22.6$ (see \citealp{2003MNRAS.343.1348P}). The conservative 
15-$\mu$m cut is adopted in S2 because in this field there are uncertainties 
in the 15-$\mu$m completeness function below this flux level.

\begin{figure}
\epsscale{0.5}
\plotone{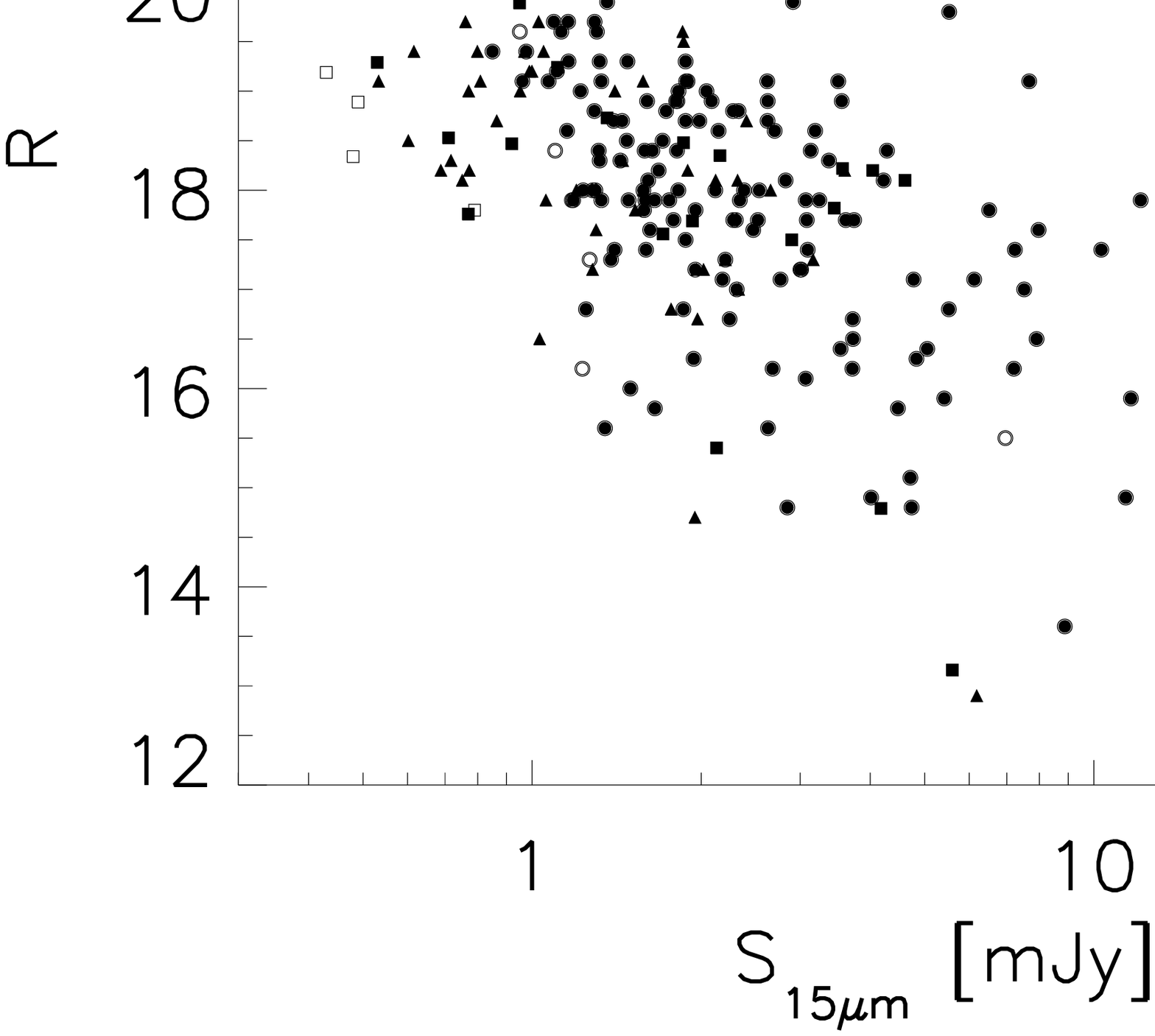}
\caption{$R$-band magnitude as a function of 15$\mu$m flux density for
all 15$\mu$m sources in the ELAIS southern fields. Sources with redshift are
shown as filled symbols, sources with likely optical counterparts but without
redshift are shown as empty symbols and sources without
optical counterparts are shown as lower limits (arrows). 
Circles stand for sources from S1$\_$rest; triangles for sources from S1$\_$5
and squares for sources from S2. The three lines represent the three
magnitude thresholds considered in this work (dotted-line: S1$\_$rest; dashed
line: S1$\_$5; dot-dashed line: S2). \label{rmag_15_fig1}}
\end{figure}

The sample of galaxies with measured redshift, after excluding objects classified as
AGNs (both type 1 and type 2) or stars on the basis of their spectra, consists of 161 galaxies:
101 in S1$\_$rest, 48 in S1$\_5$ and 12 in S2. Following \cite{fabio03} and \cite{2003MNRAS.343.1348P}, AGNs have been separated from
galaxies  according to classical diagnostic diagrams and information on the
nature of galaxy objects has been provided following the classification scheme of
\cite{1999ApJS..122...51D}. This scheme is based primarily on two lines,
[OII]$\lambda$3727 in emission and H$\delta$ in absorption, which are good
indicator of (respectively) current and recent star formation. Our sample
of 161 galaxies include 40 e(a) (spectra of dust-enshrouded starburst
galaxies), 13 e(b) (spectra with very strong emission lines), 72 e(c) (spectra
typical of spirals), 32 k(e) (spectra with signs of at least one emission
line) and 4 k (spectra of elliptical-like objects) galaxies.   
 In Figure \ref{rmag_15_fig1} we show the magnitude vs. 15-$\mu$m flux distribution
of the whole S1$\_$rest+S1$\_$5+S2 sample of galaxies (objects with spectroscopic
identification are shown with filled symbols). 

\begin{figure}
\epsscale{0.6}
\plotone{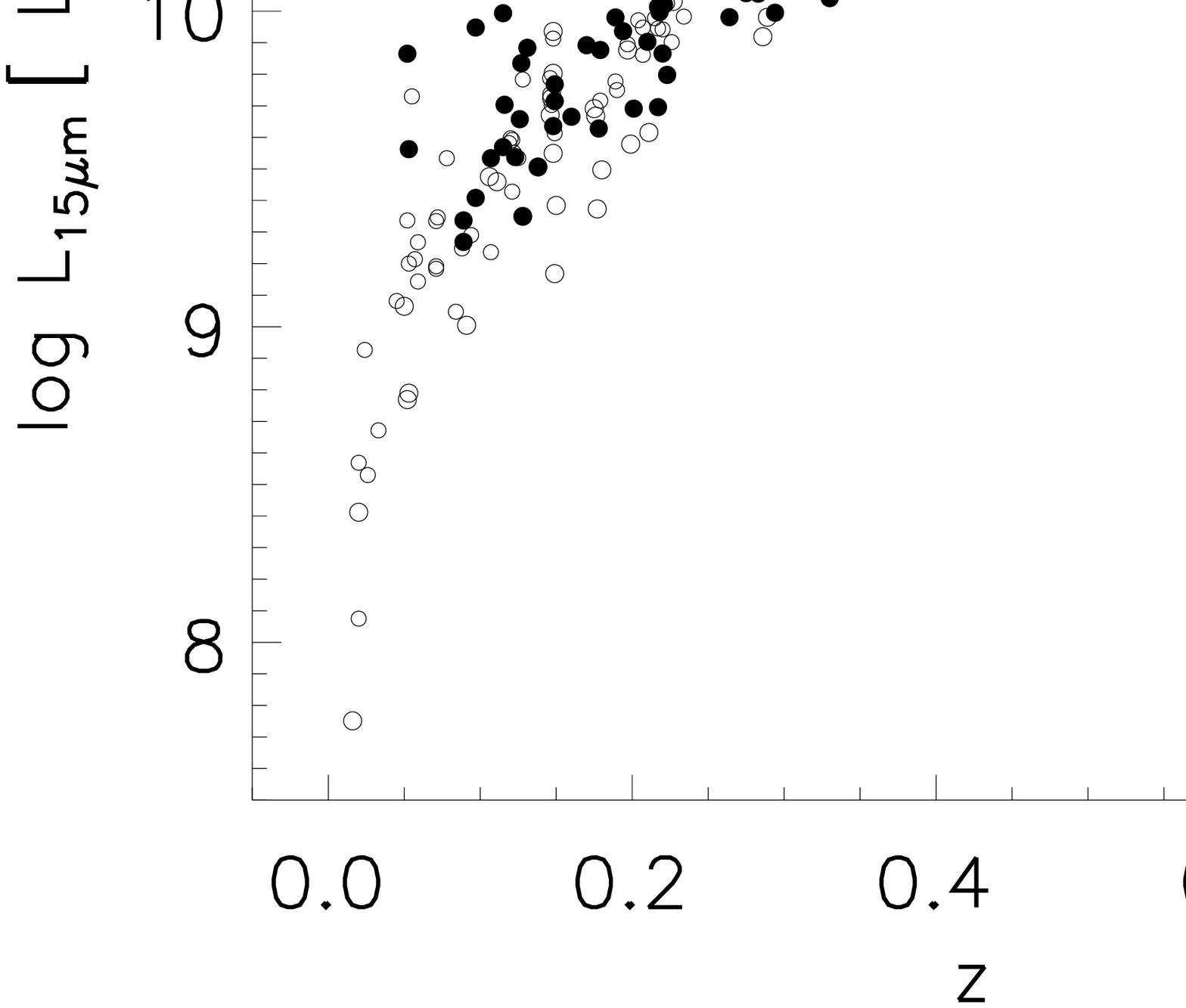}
\caption{Rest-frame 15-${\mu}$m luminosity as a function of redshift for
galaxies in the spectroscopic sample. Empty circles represent objects
with $\log(L_{15{\mu}m}/L_{\rm R})<{-0.4}$ (defined as normal galaxies in this
work); filled circles represent objects with $\log (L_{15{\mu}m}/L_{\rm R})>{-0.4}$
(defined as starburst galaxies in this work). \label{l15_z_fig}}
\end{figure}

In our estimates of the luminosity function and its evolution (Section
\ref{fdl_sec}) we apply an additional restriction, selecting only galaxies
with redshift $z\le$0.4. As shown in
Figure \ref{l15_z_fig}, in fact, above this redshift threshold the
redshift distribution of galaxies is poorly sampled. \\
The final total sample of galaxies 
considered for our LF determination
consists of 150 objects.

\placefigure{rmag_15_fig1}
\section{The K-correction}
\label{kcorr_sec}

To derive the 15-$\mu$m luminosity function from our sample of
galaxies we need to relate the observed flux $S$ (and redshift) to the
rest-frame 15-$\mu$m and $R$-band luminosities. This requires the
knowledge of the K-correction in the Mid-IR and in the $R$-band.

Models of infrared emitting populations (\citealp{2001ApJ...549..745R};
\citealp{2001A&A...378....1F}; \citealp{2002A&A...384..848E}; \citealp{2003ApJ...587...90X}) have shown that
the IR counts can be explained with three main populations. These
populations are characterised by their SED and their evolutionary
properties: the normal spiral galaxies, the starburst
galaxies and galaxies powered by an active galactic nucleus (AGNs). The
three populations can be separated on the basis of their SED in the
Mid-IR/Far-IR band (see \citealp{1989MNRAS.238..523R}; \citealp{2003ApJ...587...90X}).

\begin{figure}
\epsscale{0.7}
\plotone{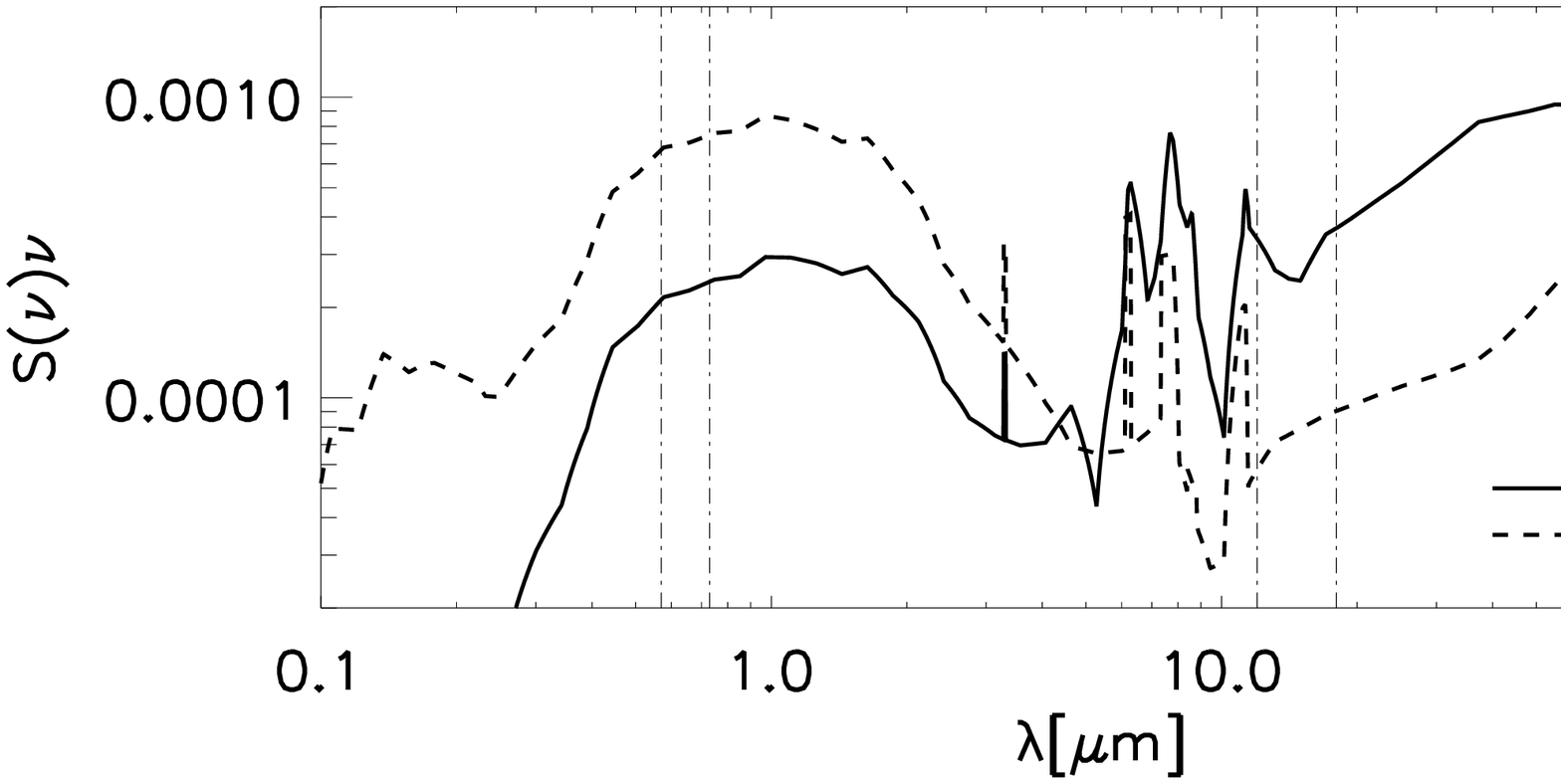}
\plotone{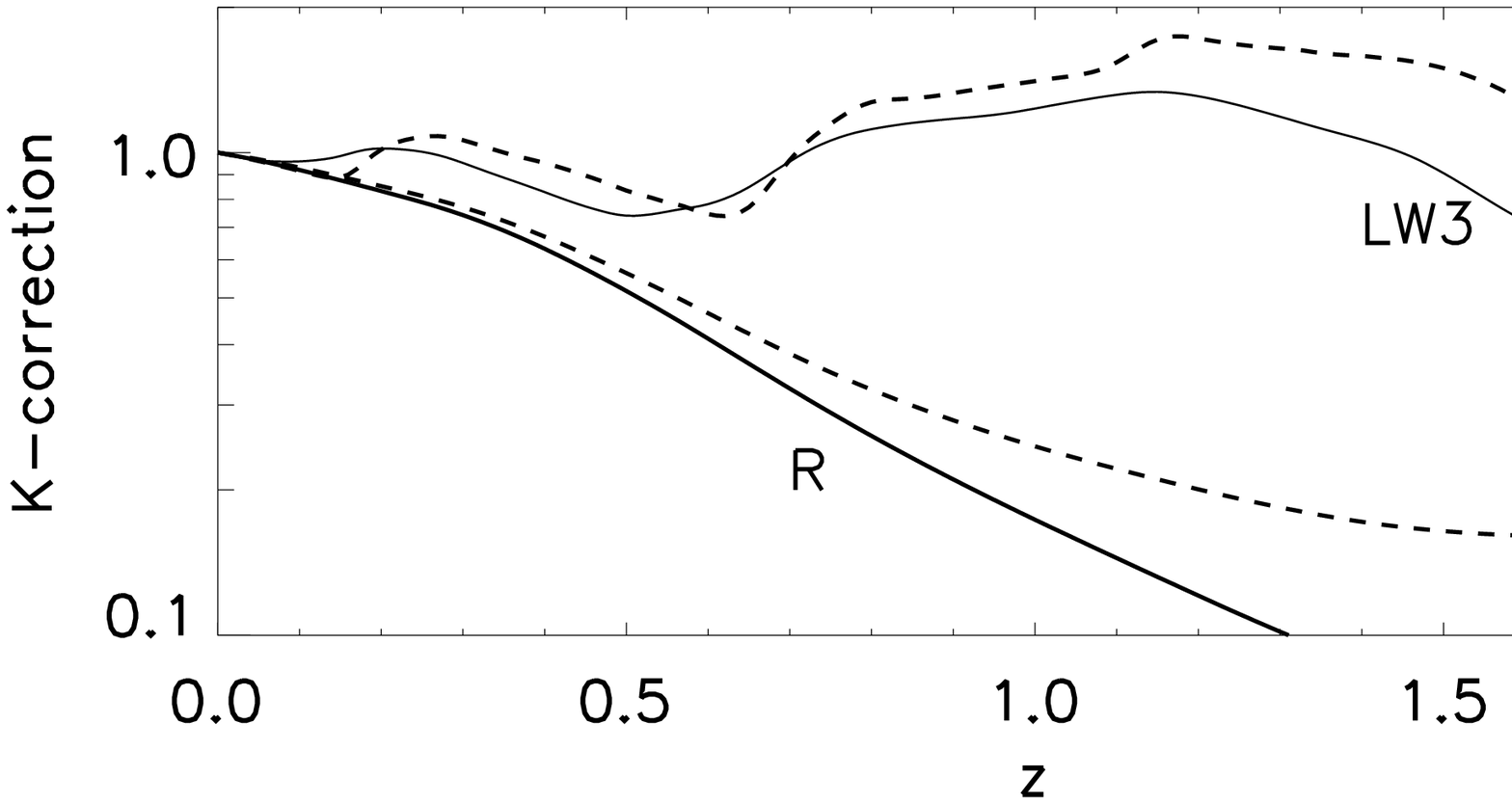}
\caption{{\it Top}: Spectral energy distributions (SED) adopted
for our galaxies. The dashed-line corresponds to M51, a low luminosity
inactive spiral galaxy, while the solid-line corresponds to M82, considered as
the star-forming galaxy prototype. The SEDs have been taken from the GRASIL
code (\cite{1998ApJ...509..103S}). In the range from 5 to 18
$\mu$m the SED for M82 is the observed ISOCAM CVF spectrum. The
vertical dot-dashed lines correspond to the $R$-band and LW3 filter
trasmissions at the half-maximum-transmission values. 
{\it Bottom}: The K-correction as a function redshift in the LW3 and $R$-band
filters for M51 (dashed-lines) and M82 (solid-lines). \label{sed_fig}}

\end{figure}

We have assumed M51 and M82 as prototypes of the normal and the
starburst populations respectively (the work on AGNs is presented in
\citealp{2002MNRAS.332L..11M}; Matute et al. in preparation). M82 is a local moderate-starburst galaxy
($L_{IR}\sim{10^{10.6}L_{\odot}}$, \citealp{2002A&A...384..848E}). We did not
consider the SEDs of a more active galaxy (like Arp220,
$L_{IR}\gsimeq{10^{11}L_{\odot}}$), since from the identifications of
ELAIS \citep{fabio03} and HDF-N sources \citep{2002A&A...384..848E},
such extreme galaxies are rare and tend to appear only at high redshift
$(z\gsimeq0.8$). The model SEDs of both M51 and M82 have been taken
from the GRASIL model output \citep{1998ApJ...509..103S} except for the
Mid-IR (5-18 $\mu$m) spectrum of M82, for which the observed ISOCAM
CVF ({\it Circular Variable Filter}) spectrum \citep{2003A&A...399..833F} has been
adopted. The two reference SEDs (normalized to the bolometric luminosity from
0.1 to 1000 $\mu$m) and the corresponding K-corrections in the
Mid-IR filter (ISOCAM LW3 filter) and in the $R$-band filter
($R$ Cousin) are shown in Figure \ref{sed_fig}. The normal spiral galaxy has
relatively high Far-IR over Mid-IR emission, and has a bolometric
luminosity dominated by the optical region of the spectrum. The active
galaxy has a bolometric luminosity
dominated by the IR emission, as expected in presence of star-formation
obscured by larger amount of dust. As shown in Figure \ref{sed_fig} (bottom),
the two galaxies have similar K-corrections in the LW3 and
$R$-band filters, at least up to $z{\sim}1$.

\begin{figure}
\epsscale{0.6}
\plotone{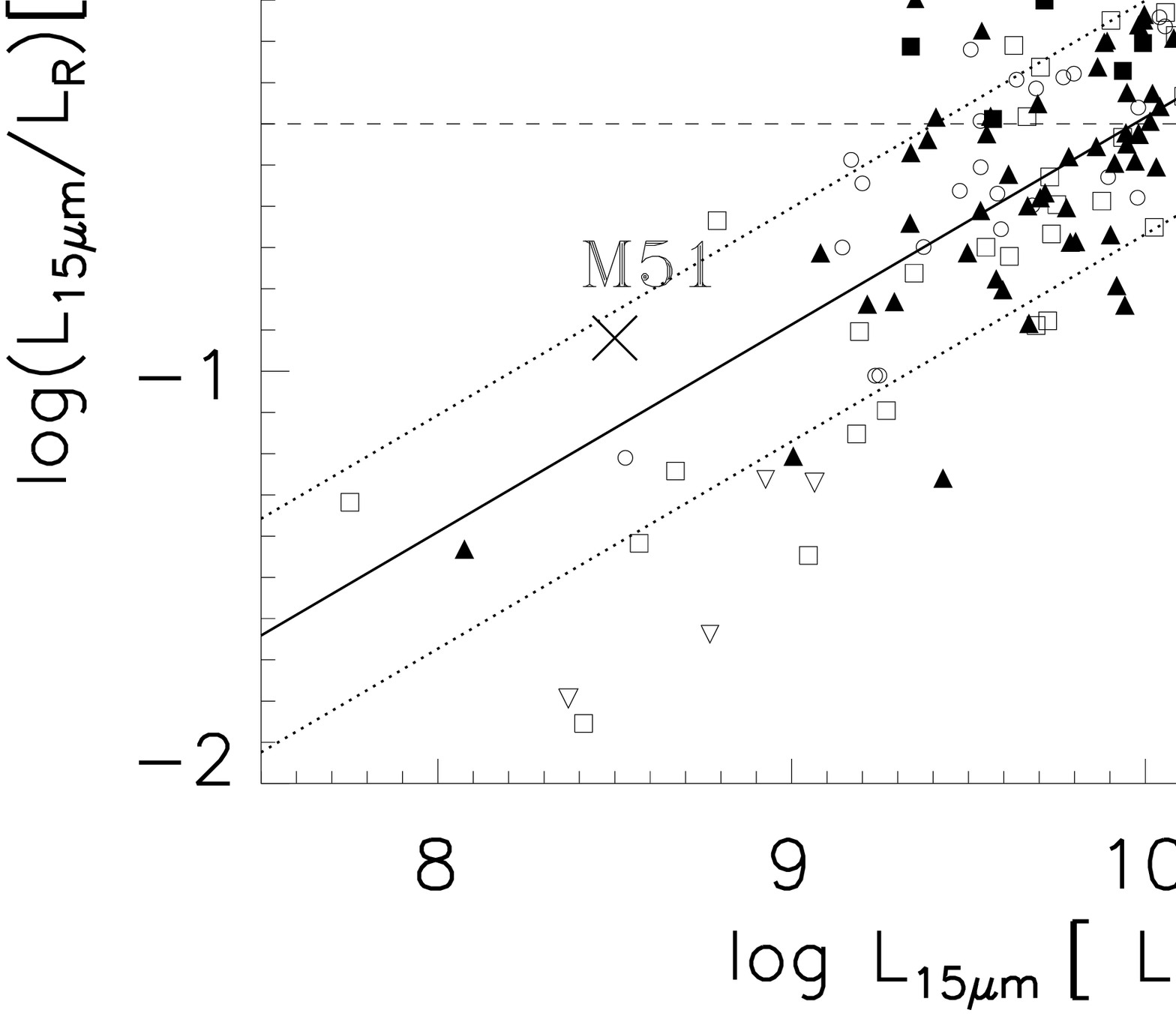}
\caption{The $L_{15{\mu}m}/L_{\rm R}$ ratio as a function of
$L_{15{\mu}m}$ for galaxies in the spectroscopic sample. Different
symbols correspond to the different spectroscopic classes (see \citealp{fabio03}). The large diagonal crosses
represent the values for M51 and M82. \label{l15_lopt_fig}}
\end{figure}

Since in our sample we do not have the Mid/Far-IR colours, we have used the Mid-IR to
optical luminosity ratio to tentatively associate a representative SED (M51 or
M82) to each galaxy. As discussed by \cite{fabio03}, galaxies selected in the Mid-IR
band show a well defined relation between the ratio of mid-infrared to optical
luminosities ($L_{15{\mu}m}/L_{\rm R}$) and the mid-infrared luminosities. This relation, although with some
significant scatter, appears to hold, over about three orders of magnitude in
$L_{15{\mu}m}$, from the bright fluxes sampled by the IRAS sources
\citep{1993ApJS...89....1R} to the faintest ISO fluxes sampled in the HDF-S
and HDF-N fields, with more infrared luminous galaxies having on average
larger $L_{15{\mu}m}/L_{\rm R}$ ratios. In first approximation, as suggested
by the average SEDs shown in Figure \ref{sed_fig}, the ratio
$L_{15{\mu}m}/L_{\rm R}$ can be interpreted as an indication of the relative importance between
bursting and more quiescent emission (see also \citealp{MRR03}).
In Figure \ref{l15_lopt_fig} we plot the rest-frame $L_{15{\mu}m}/L_{\rm R}$
versus $L_{15{\mu}m}$ for our galaxies (161 objects), where $L_{15{\mu}m}$ and $L_{\rm R}$ are the luminosities
(${\nu}L_{\nu}$) at 15-$\mu$m and in the $R$-band, respectively. The majority of
our galaxies lie between the ratio values found for M51 and M82. On the basis
of the plot, we have assumed $\log(L_{15{\mu}m}/L_{\rm R}){\sim}{-0.4}$ as the nominal separation
between the normal and the starburst galaxy populations (dot-dashed
line, see Section \ref{llf_section} for a discussion). 
This leads to define subsamples of 81 spiral galaxies (80 at $z\le$0.4)
and 80 (70 at $z\le$0.4) starburst galaxies. We did not use the spectroscopic classification to separate the
two populations, since dust can significantly depress the emission
features, which are mainly used to classify starburst galaxies, thus possibly
leading to misleading classifications (see \citealp{2000ApJ...537L..85R};
\citealp{2001A&A...378....1F}). This is evident also from
Figure \ref{l15_lopt_fig}, where the different spectroscopic classes are not
clearly segregated in different regions of the plot. In any case, it is
reassuring that almost all the objects (12/13)
spectroscopically classified as e(b) galaxies (spectra with very strong emission lines)
do indeed have $\log(L_{15{\mu}m}/L_{\rm R})\gsimeq{-0.4}$ and all the objects
(4/4) classified as k galaxies (elliptical-like spectra) have $\log(L_{15{\mu}m}/L_{\rm R})\lsimeq{-0.4}$.

A least square fitting procedure applied to the data leads to the empirical
relation:
\begin{equation}
\label{l15_lopt_eq}
\log(L_{15{\mu}m}/L_{\rm R})=0.50\;\log L_{15{\mu}m}-5.4
\end{equation}
with a dispersion of $\sim$0.28 dex. The best-fit and its $1\sigma$ bounds  are
shown in the figure as solid and dotted-lines,
respectively. This relation will be discussed in detail in the next section.

In Figure \ref{l15_z_fig} we show the $L_{15{\mu}m}-z$ diagram. The
normal and the starburst galaxies, as defined above, are represented by empty and filled
circles, respectively. The luminosity of each galaxy has been computed
using the K-correction appropriate to its classification. Considering the
galaxies up to $z=0.4$, the median redshift of the whole sample (spiral plus
starburst galaxies) is $z_{\rm med}{\simeq}$0.18, while the
median luminosity is $L_{15{\mu}m}{\simeq}10^{9.8}L_{\odot}$. The
majority of the normal galaxies lie at relatively low redshift
($z_{\rm med}{\simeq}$0.14) and luminosity
($L_{15{\mu}m}{\simeq}10^{9.5}L_{\odot}$), while the active
population is characterized by higher median values:
$z_{\rm med}{\simeq}$0.23 and $L_{15{\mu}m}{\simeq}10^{10}L_{\odot}$.

\placefigure{sed_fig}

\section{The Mid-IR Luminosity Function}
\label{fdl_sec}

\subsection{Estimator and Selection Effects}

The luminosity function and its evolution have been estimated using
the parametric, unbinned, maximum likelihood method described in \cite{1983ApJ...269...35M}.  We consider three different
selection effects affecting our data: the 15-$\mu$m, the optical
$R$-band and the spectroscopic limits.

The 15-$\mu$m selection effect has been corrected for by
weighting each data pair ($z$,$L_{15{\mu}m}$) by its 15-$\mu$m
effective area ($\Omega(z,L_{15{\mu}m})=\Omega(S)$). The
completeness functions for S1 and S1$\_$5 are given in Table 1 of \cite{fabio03}; for S2, we consider only fluxes greater than 1 mJy
with an associated completeness of 93 \% for $1 < S < 2$ mJy and 100 \% for $S \ge 2 $mJy (\citealp{2003MNRAS.343.1348P}).

The optical limits, reported in Section \ref{spectra_sec}, have been
taken into account by introducing a function $\Theta(z,L_{15{\mu}m})$
which represents the probability that a source, with a given 15-$\mu$m
luminosity $L_{15{\mu}m}$ and redshift $z$, had an $R$-magnitude within
the limits of the sample. To calculate the probability of any optical
luminosity ($L_{\rm R}$) given a specific data pair ($z$,$L_{15{\mu}m}$)
we use the relation between the 15$\mu$m and the optical luminosities
together with its dispersion, assumed to be Gaussian (Eq. \ref{l15_lopt_eq}). The
$R$-band K-correction described in the previous section has been used to relate the
optical rest-frame luminosity to the observed magnitude.

The spectroscopic selection has been considered by weighting each
triplet $(z,L_{15{\mu}m},L_{\rm R})$ by the spectroscopic completeness
 in the corresponding optical interval. This correction is not
significant because of the high spectroscopic completeness in
the considered optical intervals (${\gsimeq}95$ \% , see Section
\ref{spectra_sec}).

Following \cite{1983ApJ...269...35M} the function to be minimized can be
written as $S=-2\ln{\cal L}$,  where ${\cal L}$ is  the likelihood function:
\begin{eqnarray}
\label{like_eq}
S&=&-2\sum_{i=1}^{N}{\ln\phi(z_{i},L_{i})}\\\nonumber
 &+& 2\int\int\phi(z,L)\,\Omega(z,L)\,\Theta(z,L)\;\frac{dV}{dz}\;dz\;d{\log}L
\end{eqnarray}
where $L$ is the luminosity at 15 $\mu$m ($L_{15{\mu}m}$), $N$ is the total number of sources in the three samples, $\Omega(z,L)$
is the available area of the sky for an object with luminosity $L$ at redshift
$z$, $\Theta(z,L)$ is the optical correction factor, $(\frac{dV}{dz})$ the differential volume
 element and $\phi(z_{i},L_{i})$ is the luminosity
function.\\
To optimally combine the information from the three different samples 
(S1$\_$rest, S1$\_$5 and S2), we follow the formalism described in \cite{1980ApJ...235..694A}. Each factor of the double integral of Eq. \ref{like_eq} is the
sum of three terms:
\begin{eqnarray}
\Omega(z,L)\,\Theta(z,L) & = & A_{S1}\,C_{S1}(z,L)\,\Theta_{S1}(z,L)  \\
                         & + & A_{S15}\,C_{S15}(z,L)\,\Theta_{S15}(z,L)\nonumber\\
                         & + & A_{S2}\,C_{S2}(z,L)\,\Theta_{S2}(z,L)\nonumber 
\end{eqnarray}
where $A_{S1}$, $A_{S15}$ and $A_{S2}$ are the areas of the three fields (3.55, 0.55 and 0.12 deg$^2$ for S1, S1$\_5$ and S2 respectively), 
$C_{S1}$, $C_{S15}$,$C_{S2}$ are the
three completeness functions and $\Theta_{S1}$, $\Theta_{S15}$ and
$\Theta_{S2}$ are the three optical factors.

Since we do not have enough data to assess different parametric forms,
we decided to parameterize the luminosity function $\phi(L)$ using the form
suggested by \cite{1990MNRAS.242..318S} as a good description of local 60-$\mu$m
luminosity function of IRAS galaxies. By using a large sample of
sources ($\sim$2800 objects) \cite{1990MNRAS.242..318S} found
that a Schechter function was too narrow to describe IR selected
sources and a better fit could be achieved using the function:

\begin{eqnarray}
\phi(L)&=&\frac{dN(L,z=0)}{\;dz\;d{\log}L}\\
       &=&\phi^{\star}\left(\frac{L}{L_\star}\right)^{1-\alpha}\exp\left[-\frac{1}{2{\sigma}^2}{\log}^2_{10}\left(1+\left(\frac{L}{L_\star}\right)\right)\right]\nonumber
\end{eqnarray}

\placefigure{l15_lopt_fig}

\placefigure{l15_z_fig}

In our likelihood analysis, we have searched for the best fitting
parameters of the local luminosity function and simultaneously tried
to constrain the evolution of the Mid-IR galaxies to reproduce the
observed distribution of our data in the ($z$,$L_{15{\mu}m}$) plane.

As suggested by \cite{2001A&A...378....1F}, the shape of the observed
source counts (Euclidean from IRAS to a few mJy, followed by a sharp
upturn at fainter fluxes) favors the hypothesis of strong
evolution for only a fraction of the whole population, with the
remaining galaxies giving rise to the Euclidean behavior. We have
thus constrained our model by assuming that the spiral population
does not evolve, while allowing the active galaxies (starbursts) to
evolve both in density and in luminosity, according to
\begin{equation}
\phi(L,z)=g(z)\phi\left(L/f(z),0\right)
\end{equation}
parameterizing the two evolutions with two power-laws:
$g(z)=(1+z)^{k_d}$ and $f(z)=(1+z)^{k_l}$.

Since the likelihood method determines the shape and evolution of the
luminosity function, but not the overall normalization, we have normalized the
two luminosity functions (for normal spiral and active galaxies), by requiring
agreement between the predicted and the observed total number of sources for each class of objects.

\subsection{Results} 
\label{result_sec}

\begin{figure}
\epsscale{0.5}
\plotone{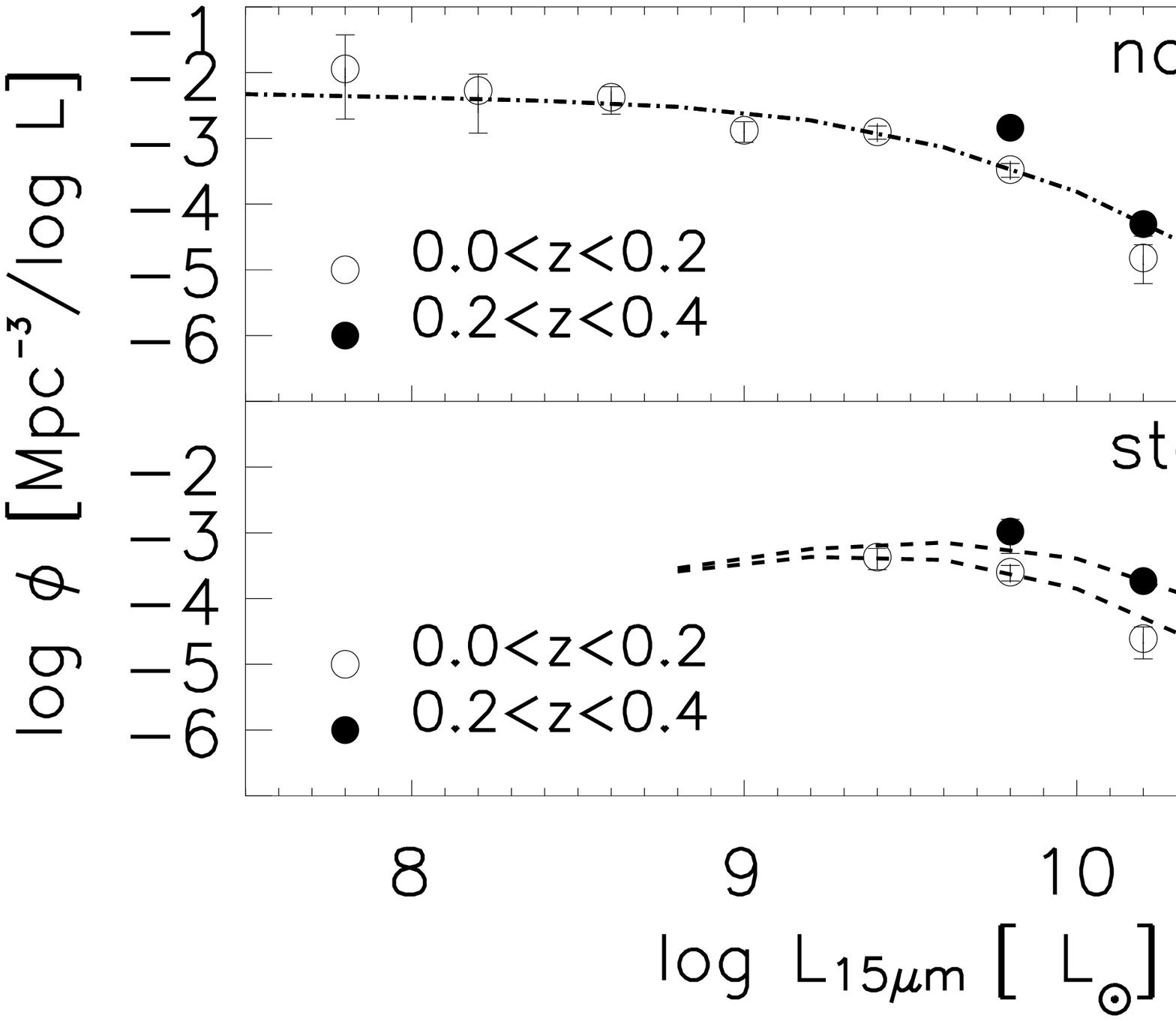}
\plotone{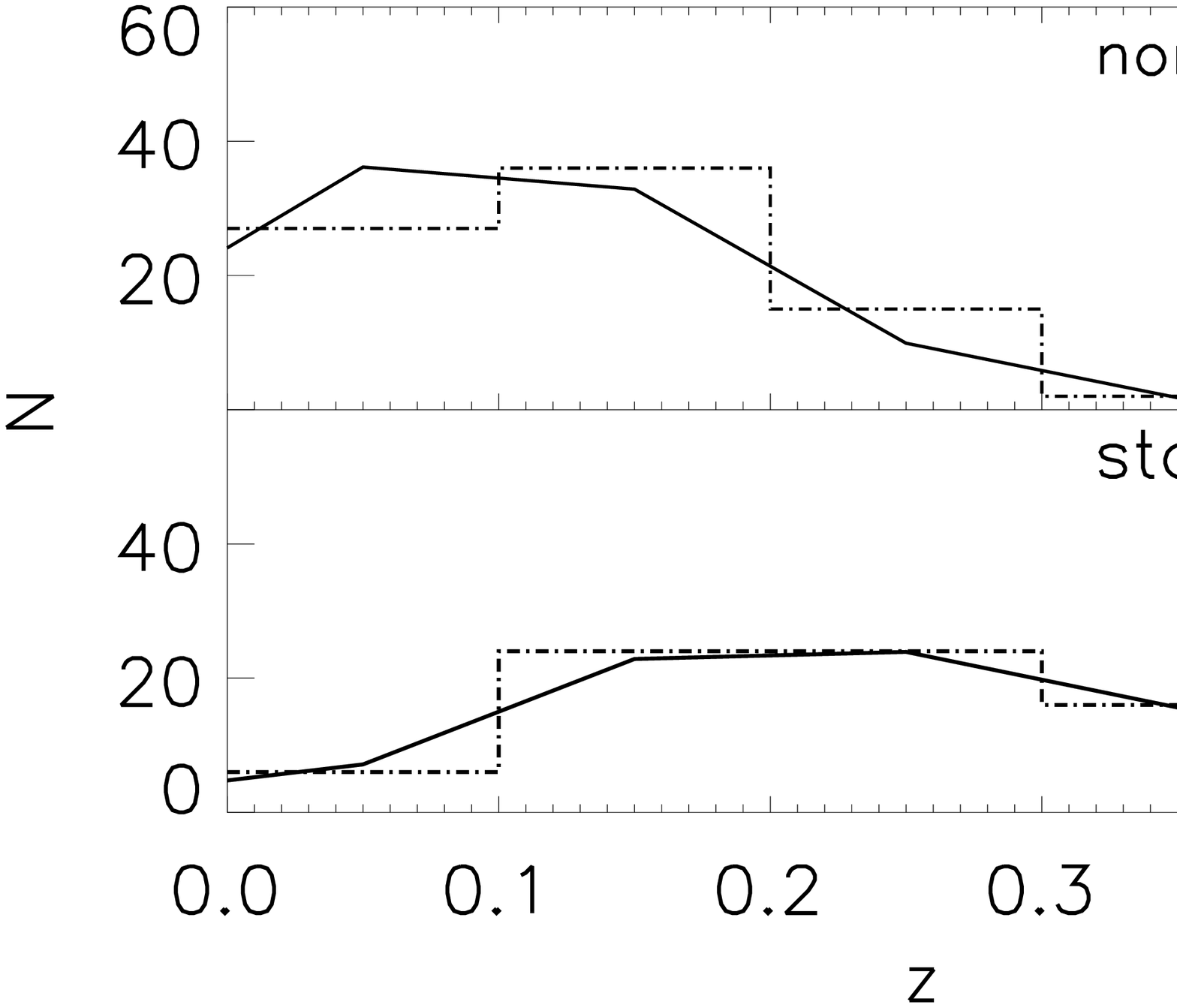}
\caption{{\it Top}: The rest-frame 15-$\mu$m luminosity functions for the
two galaxy classes from our survey in the two redshift bins $0.0<z \le 0.2$ and
$0.2<z \le 0.4$. The empty circles represent the $1/V_{max}$ determination
in the redshift interval 0.0 -- 0.2, while the filled circles are in the interval 
0.2 -- 0.4. {\it upper plot}: normal galaxy population; {\it lower plot}:
starburst population. {\it Bottom}: Comparison between the
observed (dot-dashed histogram) and the predicted redshift
distributions (solid-line) for the two galaxy classes. {\it upper plot}: 
normal galaxy population; {\it lower plot}: starburst population. \label{fdl_spi_sta_fig}}
\end{figure}

\placefigure{fdl_spi_sta_fig}

Figures \ref{fdl_spi_sta_fig}a,b show the results of our ML best fit to the
luminosity function of spiral and starburst populations in two
different redshift bins: $0.0<z\le0.2$ ($z_{\rm mean}{\sim}0.12$) and
$0.2<z\le0.4$ ($z_{\rm mean}{\sim}0.27$). The plotted data points correspond
to the space densities of the observed sources computed independently with the
$1/V_{max}$ formalism (\citealp{1968ApJ...151..393S}; \citealp{1976ApJ...207..700F}).

The best-fitting parameters for the local luminosity functions of the
two populations are reported in Table \ref{fdl_tab}. The quoted errors
correspond to the 1$\sigma$ confidence limit for each parameter
calculated while allowing all the other parameters to vary ($\Delta{\cal L}$=1, see
\citealp{1976ApJ...208..177L}). While for the starburst population we have allowed the evolution parameters to
vary, as said before, we have assumed no evolution for the spiral component. For this reason, no error is reported for its evolution
parameters. Support to this hypothesis is given by the $V/V_{max}$
  test \citep{1968ApJ...151..393S}: in fact, under the hypothesis of no
  evolution we find $V/V_{max}=0.55\pm0.03$ for the normal spiral population and
  $V/V_{max}=0.64\pm0.03$ for the starburst population ($\gsimeq4\sigma$
  evidence of evolution). The latter value becomes $0.52\pm0.03$ assuming the evolution rates reported in Table
\ref{fdl_tab}.

Because of the relatively small number of objects in each population,
the parameters derived from our maximum likelihood procedure are not
very well constrained. This is true in particular for the evolutionary rates
of the starburst population. In fact, although our data indicate a strong
evolution for this population, the uncertainties 
on the evolutionary parameters are large due to the limited redshift 
interval covered by our surveys. 
For this reason, in order to better test the evolution, we have considered 
also other observables like source counts and redshift distributions at higher
$z$, as will be discussed in Section \ref{evolution_sec}.

As shown in Figure \ref{fdl_spi_sta_fig}a, the two populations sample different
regions of the luminosity-redshift plane. In particular, the normal spiral population, which mainly comprises low redshift and low luminosity
galaxies, samples well the faint end of the luminosity function,
allowing an accurate determination of the $\alpha$ slope ($\sim$20 \%
uncertainty, see Table \ref{fdl_tab}). On the other hand, the starburst
population samples well the high luminosity, moderate-high redshift regions. This allows the
knee of the luminosity function to be better sampled and the $L_{\star}$ and
$\sigma$ parameters to be determined quite accurately. On the other hand,
since we have not starburst galaxies at low luminosities (see
Fig. \ref{fdl_spi_sta_fig}), our data do not allow to constrain, for this
population, the $\alpha$
slope, which has been fixed to 0.0 as reported in Table \ref{fdl_tab}.

To test the consistency between our observed $z,L_{15{\mu}m}$
distribution and that predicted from our parametric model we have
performed a 2-dimensional Kolmogorov-Smirnov test (2D-KS, see
\citealp{1983MNRAS.202..615P}; \citealp{1987MNRAS.225..155F}). 
The 2D-KS test gives  $\ge$10 \% probability that the observed data are
randomly sampled from the distributions predicted by the fitted LF. 


\placetable{fdl_tab}

\subsection{The unidentified objects}
\label{unid_sec}

In our maximum likelihood procedure, we have considered only objects with
$R$ mag brighter than our adopted limits and with $z\lsimeq$0.4 (150 objects). We have then
corrected the remaining redshift incompleteness inside the considered magnitude range
by applying weights to each galaxy with spectroscopic redshift. In this way,
we have assumed that the objects with no $z$ (but within the $R$ mag limits) have the same properties 
(i.e. follow the same $L_{15{\mu}m}/L_{R}$ vs. $L_{15{\mu}m}$ relation) as the
spectroscopically identified objects with similar $R$ magnitude.
Since the redshift completeness in the considered magnitude intervals is always very high ($\gsimeq$95\%), we are confident that the uncertainties
introduced are negligible.

A more delicate task is to deal with the 15-$\mu$m sources which have an $R$ magnitude fainter than the selection
limits or are unidentified (125 objects, $\sim$36 \% of
the non-stellar sample, see \citealp{fabio03} and Fig. \ref{rmag_15_fig1}). These sources could be either higher
redshift sources, or optically
less luminous galaxies at redshift similar to those of the spectroscopically
identified sample. In the first case, they are expected to follow the 
$L_{15{\mu}m}/L_{R} - L_{15{\mu}m}$ relation as the other galaxies, 
while this would not the case for the second hypothesis. 

\begin{figure}
\epsscale{0.55}
\plotone{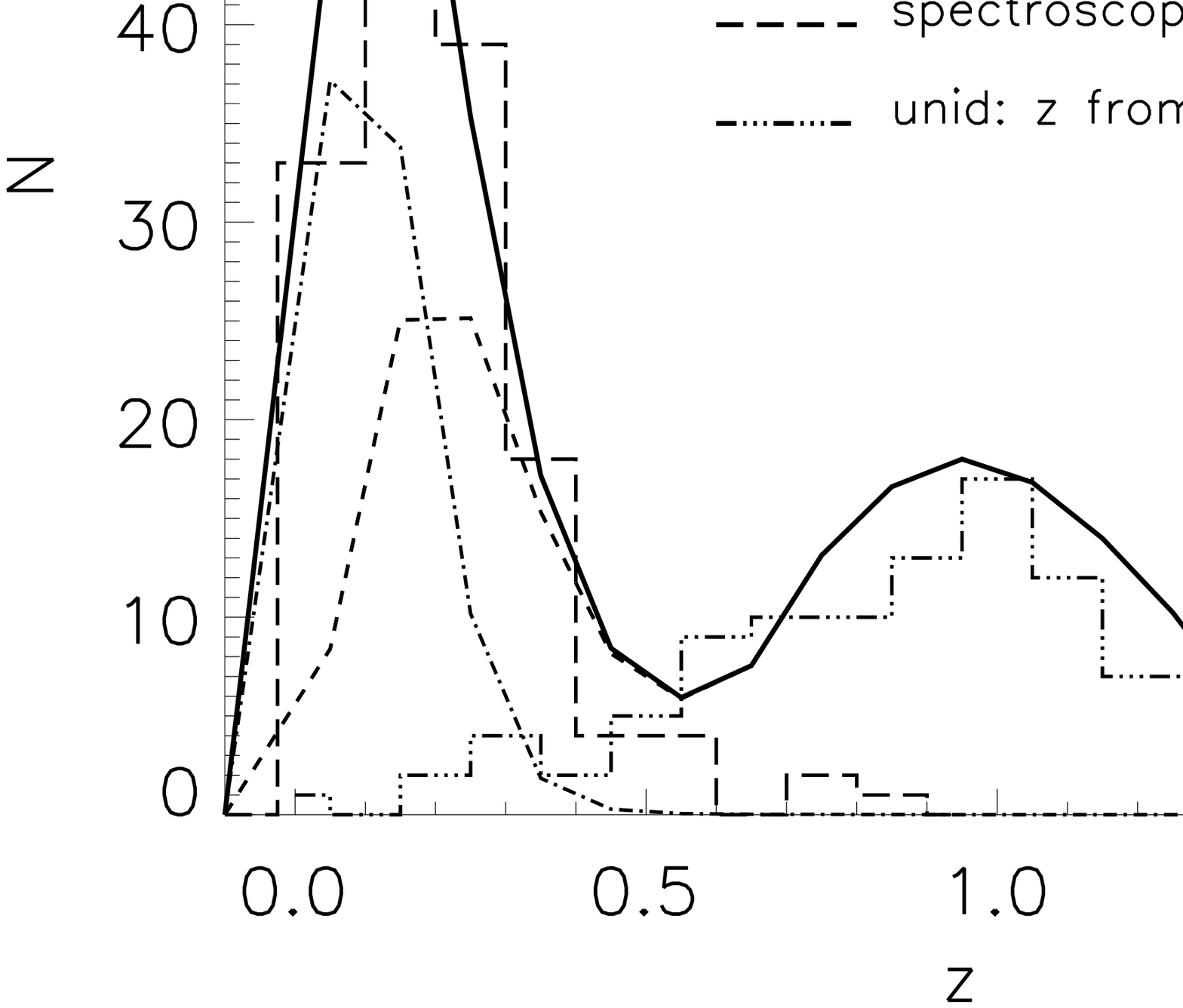}
\caption{Comparison between the observed redshift distribution for the spectroscopically
  identified objects in the total S1$\_$rest+S1$\_$5+S2 sample (long-dashed histogram) and
  the model predictions (solid-line). The contributions of the normal spiral and
  starburst components are shown as dot-dashed and dashed lines
  respectively. The estimated distribution of the
  unidentified objects as given in \cite{fabio03} is also
  shown as dot-dot-dot-dashed line. \label{z_prev_elais_fig}}
\end{figure}

Following \cite{fabio03}, the first scenario seems more probable. The considerations
discussed in \cite{fabio03} for this choice were mainly two: first the $L_{15{\mu}m}/L_{R}$
 vs. $L_{15{\mu}m}$ seems to be a valid relation for all the Mid-IR surveys, from the ISOCAM ultra-deep
 to the local IRAS surveys; second, by estimating the redshift of the unidentified
 objects on the basis of the observed $\log{z}-R$ relation, the computed
 $L_{15{\mu}m}$ and $L_{R}$ appear well consistent with the assumed
 relation. In Figure \ref{z_prev_elais_fig}, the observed redshift distribution of 
the spectroscopically identified objects and the estimated
distribution (on the basis of the $\log{z}-R$ relation) of the
unidentified sources is compared with the distribution predicted by our model (by
extrapolating the results to high-$z$, see Section
\ref{evolution_sec}).
In Figure \ref{z_prev_elais_fig}, the agreement between the model predictions and the observed data (including
the estimated distribution) is quite impressive. In particular, the
estimated $z$-distribution of the
unidentified sources is expected to fill exactly the secondary peak
predicted by the model.

A further evidence supporting the hypothesis that most of the unidentified
sources should belong to the same population as the identified
  sources, but with higher redshift ($z=0.5-1.5$) is supplied by the
  photometric redshift technique. The photometric redshifts have been
  estimated and presented in the Final
  Band-merged ELAIS Catalogue (\cite{MRR03}). The final Band-merged
  catalogue contains 1636 15-$\mu$m sources, 136 of which with $R
  \gsimeq 20.5$ (optical limit of our larger sample, S1$\_$rest, see
  Section \ref{spectra_sec}) and enough optical data to determine
  photometric redshifts. The resulting photometric redshift
  distribution for all the ELAIS fields is in agreement with our
  hypothesis, showing that most of the 15-$\mu$m sources
  with associated faint optical
  galaxies are at significantly higher redshift than sources with brighter optical counterparts ($z_{mean}=0.75$ instead of $z_{mean}=0.2$ of the present optical bright sample).

The bimodal behavior of the $z$-distribution of the 15-$\mu$m sources
shown in \ref{z_prev_elais_fig} is
caused by a combination of different causes: the shape of the LW3
  K-correction (see Fig. \ref{sed_fig}), the high evolution rates found for
  the starburst population and, finally, the
  slope of the LF for high 15-$\mu$m luminosities. The dip in the redshift
  distribution around $z\approx$0.5 is probably the cause of the gap around $R\sim$21 of the distribution of the optical counterparts of the
15-$\mu$m sources (see Fig. \ref{rmag_15_fig1} and the discussion in \citealp{fabio03}).

\placefigure{z_prev_elais_fig}
\placetable{fdl_vmax_tab}

\section{The local luminosity function}
\label{llf_section}

\begin{figure}
\epsscale{0.6}
\plotone{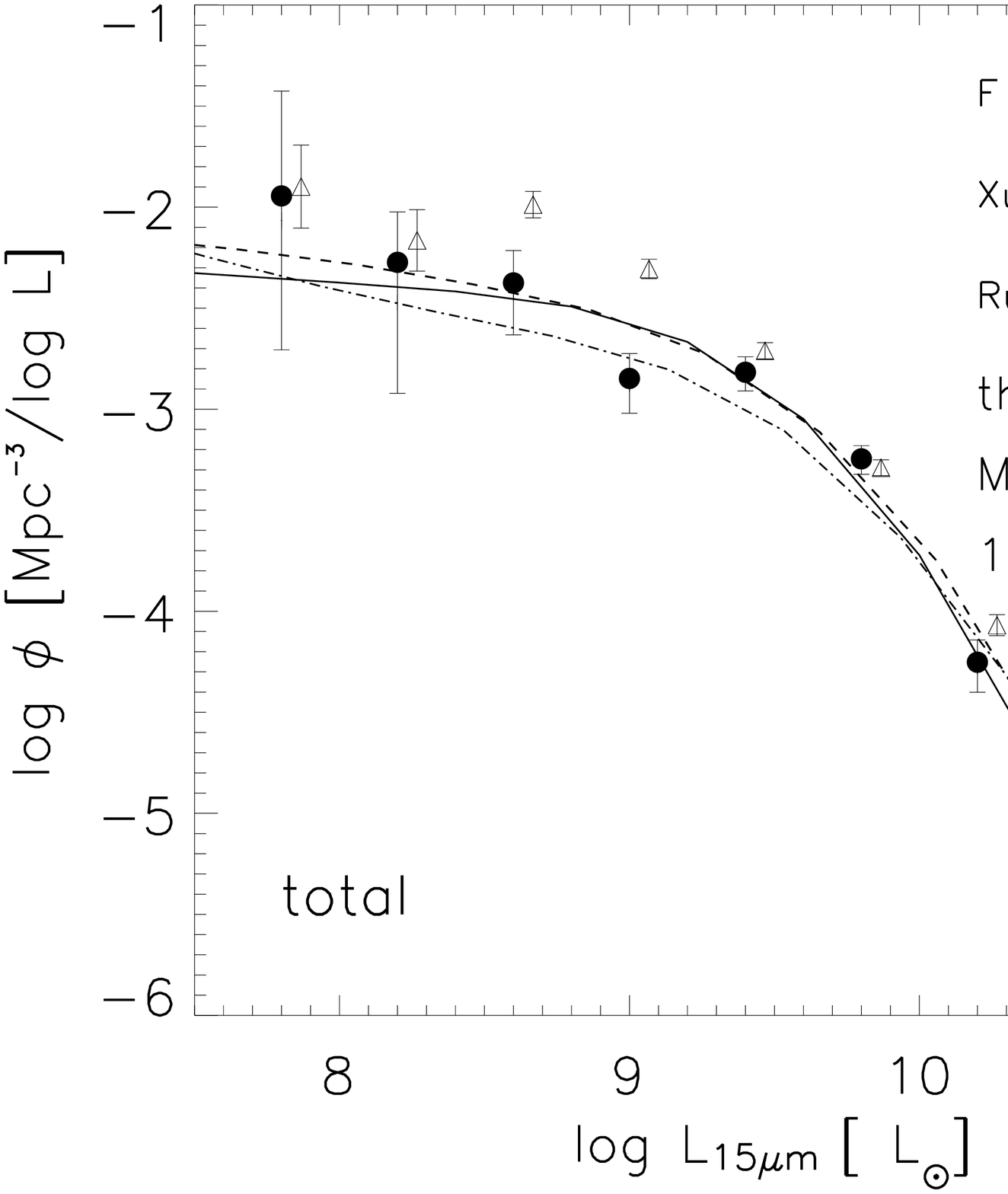}
\caption{Local 15-$\mu$m luminosity function of galaxies. The filled circles
  are the $1/V_{max}$ estimates from this work, while the solid line is the
  extrapolation to $z=0$ of the ML results. The dashed line is the 12-$\mu$m LLF from \cite{2001A&A...378....1F} (the Sey2 contribution has been subtracted using the LLF for Sey2 computed by \citealp{1993ApJS...89....1R}). The dot-dashed line is the 15-$\mu$m conversion of the
 LLF computed by Xu
  at 25-$\mu$m. Empty triangles are an estimate at 12-$\mu$m of the LLF of galaxies, based on the \cite{1993ApJS...89....1R} catalogue. The LLF computed at Mid-IR bands different from
  15-$\mu$m have been converted to 15-$\mu$m as explained in the text. \label{llf_fig}}
\end{figure}

In Figure \ref{llf_fig} the local luminosity function (LLF) of 15-${\mu}$m galaxies
(excluding AGNs) estimated in this work is compared with other determinations
derived at different Mid-IR bands and converted to 15-${\mu}$m using our SEDs
(M51 or M82, depending on galaxy type).

Our estimate of the LLF has been done by extrapolating the result
of the maximum likelihood method to $z=0$ (solid line). To check the
over-all normalization we have used the
$1/V_{max}$ formalism, `de-evolving' each galaxy according to the derived
evolution coefficients (filled circles). 
In Table \ref{fdl_vmax_tab} the $1/V_{max}$ LLF is
listed, with $L_{15{\mu}m}$ defined as ${\nu}L_{\nu}$ and bin width
$\delta\log(L_{15{\mu}m})=0.4$. The results of the two methods are well
consistent with each other and the final $V/V_{max}$
value for the total sample is $V/V_{max}$=(0.53$\pm$0.02).

The open triangles are an estimate of the LLF of all galaxies (excluding Sey1 and
Sey2) based on the 12-$\mu$m catalogue of
\cite{1993ApJS...89....1R}. We have computed this LLF using the $1/V_{max}$ method,
selecting all galaxies with $S_{12{\mu}m}>300$ mJy for which the
12-$\mu$m, optical and spectroscopic completeness are 100 \%
(see \citealp{1993ApJS...89....1R}).
The dashed-line is the LLF assumed by \cite{2001A&A...378....1F} and is 
based on the 12-$\mu$m LLF computed by \cite{1998ApJ...500..693F} and re-adapted by \cite{1998ApJ...508..576X} in the low
luminosity regime. \cite{2001A&A...378....1F}
assume that starburst galaxies contribute $\sim$10
\% of the 12-$\mu$m LLF at all luminosities. Therefore we have converted their
 12-$\mu$m LLF to 15-$\mu$m by using the M82 and M51 SEDs for $\sim$10
\% and  $\sim$90\% of the 12-$\mu$m LLF, respectively.
Moreover, since in the \cite{2001A&A...378....1F} model the active 
population includes also the Sey2 galaxies, we have obtained only the starburst contribution by 
subtracting the LLF of Sey2 computed by \cite{1993ApJS...89....1R}.
Finally, the dot-dashed line is the LLF computed by \cite{2001ApJ...562..179X}, which is
based on the 25-${\mu}$m sample of \cite{1998ApJ...501..597S}. This LLF is the sum of
an actively starforming and a normal population, defined on the basis of the IRAS
colours. 

As shown in Figure \ref{llf_fig}, the three independent estimates are reasonably
well consistent with each other. The differences in the low
luminosity regime between the \cite{1993ApJS...89....1R} and the other determinations
is possibly caused
by the effect of local inhomogeneities (particularly the Virgo super-cluster)
in the IRAS survey, as suggested by many authors (\citealp{1998ApJ...500..693F}; \citealp{1998ApJ...508..576X}). At high
luminosity, the tendency of the \cite{2001ApJ...562..179X} LLF to be higher 
than the others is probably
due to contamination from Sey2 objects, showing IR colours similar to those of
starbursts. It must be underlined the agreement between the overall normalization of
our determination (based on the ISO data) and the other estimates based on
IRAS data. This is indicative of a great accuracy both in the ISO calibration 
achieved with the {\it Lari method} (see \citealp{2001MNRAS.325.1173L}) and in the 
completeness corrections applied to our data.

\placefigure{llf_fig}

\begin{figure}
\epsscale{0.45}
\plotone{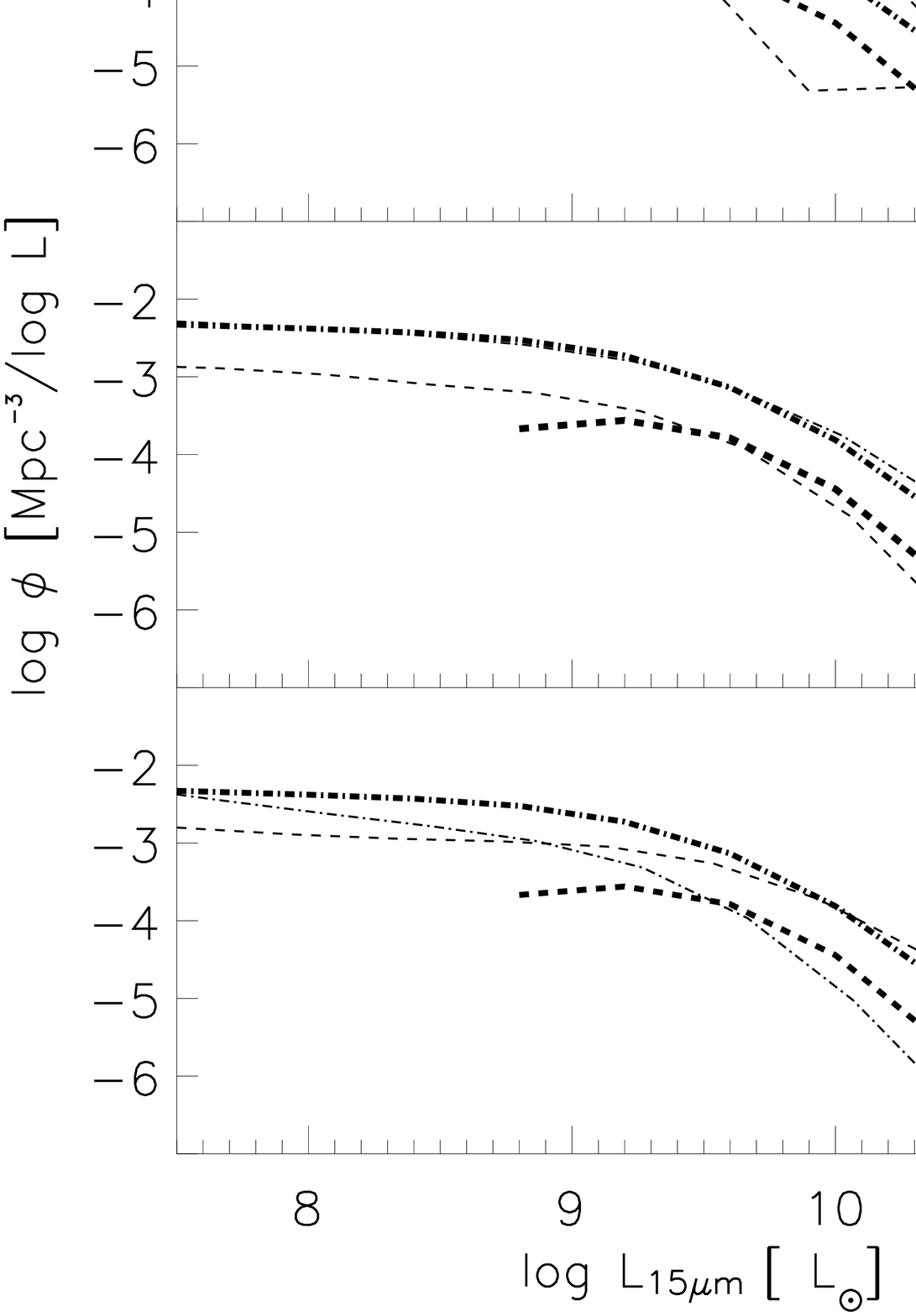}
\caption{Comparison between the LLF subdivision into starburst and normal
galaxy populations from this work and other analyses. The starburst galaxies are shown as dashed
lines while the spirals as dot-dashed lines. The determinations from this work
are thicker. {\it Top}: comparison with \cite{1993ApJS...89....1R}; {\it
  Middle}: comparison with \cite{2001A&A...378....1F} (the LLF computed by
\cite{1993ApJS...89....1R} for the Sey2
galaxies has been subtracted from the active component). {\it Bottom}: comparison with \cite{2003ApJ...587...90X}. \label{llf_pop_fig}}
\end{figure}

While the total determinations of the Mid-IR LLF
for galaxies agree so well, their subdivisions into different
populations (starburst and normal galaxies) made by different
authors do not show the same level of consistency.
In Figure \ref{llf_pop_fig} our LLF for the two galaxy populations (thick lines)
are compared to those derived by \cite{1993ApJS...89....1R},
\cite{2001A&A...378....1F} and \cite{2003ApJ...587...90X} (top, middle and
bottom panel respectively). The starburst populations are shown as dashed lines and the
spiral ones as dot-dashed lines. The \cite{1993ApJS...89....1R} 12-$\mu$m LLF
for normal galaxies (shifted to 15-$\mu$m through the M51 SED)
is almost identical to our determination for the same population.
Instead, the LLF of their $liner + starburst$ component (the latter defined as
sources with high FIR luminosity) is significantly
different from that estimated for our starburst population (basically it is much flatter than ours at $L_{15{\mu}m} \gsimeq 10^{10}~
 L_{\odot}$ and has a much higher volume density for the highest luminosity
 objects, $L_{15{\mu}m} \sim 10^{11}~
 L_{\odot}$). The \cite{2001A&A...378....1F} LLF for galaxies
are very similar to ours, both for starburst and normal spirals (also in this
case, the LLF of Sey2
computed by \cite{1993ApJS...89....1R} has been subtracted from the active component 
of \cite{2001A&A...378....1F}). The \cite{2003ApJ...587...90X}  
populations' subdivision, derived from IRAS colours of
25-$\mu$m selected sources, is totally different, almost
opposite to ours. In fact, their starburst LLF is higher than the normal galaxy one for $L > 10^9~ L_{\odot}$ 
and its high luminosity slope is rather flat,
producing a significant local contribution of starburst galaxies
even at $L > 10^{10.5}-10^{11}~ L_{\odot}$. On the contrary,
the \cite{2003ApJ...587...90X} normal galaxy LLF is lower and steeper
at high luminosities than the starburst one, and is more similar
to our determination for starbursts.
This significantly different subdivision, together with different
evolutionary schemes proposed, is probably the main cause of
the large difference in the model predictions between our
work and that of \cite{2003ApJ...587...90X} .

The populations' separation adopted in our work is based on a
physical property of galaxies, like the ratio between
Mid-IR and optical luminosity. The choice of the value of the
ratio ($R = L_{15{\mu}m}/L_{\rm R}$) chosen to divide the starburst from 
the normal spiral population is mainly based on two reasons.  
First, as previously said, it is intermediate between
the value observed for M51 (normal spiral) and that observed for M82 (starburst galaxy). 
Second, it is the value that best allows to reproduce all the observables, from the normalization of the
LLF to the redshift distributions at low- and high-$z$ and source counts at all flux levels. A
smaller fraction of the normal, non-evolving,  population ($\log R{\lsimeq}-0.6$),
while would have allowed to better model the sharp increase of the
ELAIS-S1 counts (since a lower contribution of the spiral component would have
been produced at mJy level), it would have underestimated the total local
luminosity function. On the contrary, a larger fraction of the normal population
($\log R{\gsimeq}-0.2$) would have enhanced the contribution of the quiescent component in
the sources counts, predicting a too smooth behavior with respect to the
counts shape observed in the ELAIS-S1 field.

\placefigure{llf_pop_fig}

\begin{figure}
\epsscale{0.6}
\plotone{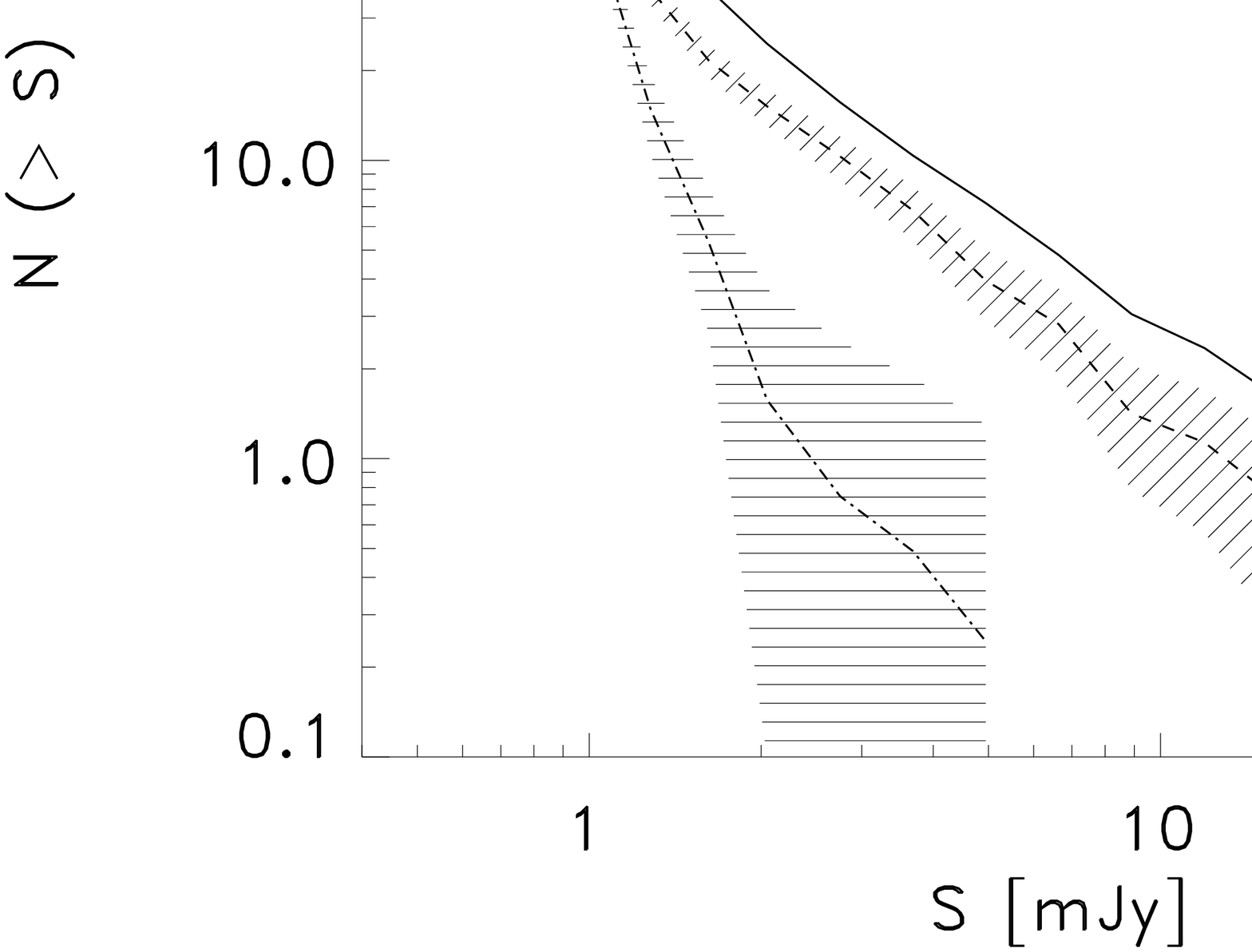}
\caption{The contribution to the total integral source counts (solid
line) from galaxies in the spectroscopic sample (dashed-line) and from
unidentified sources (dot-dashed line) in the ELAIS-S1 survey. The
areas filled with horizontal and diagonal lines represent \ the 68\% confidence regions. The confidence region for the total
counts have not been
shown for clarity. \label{counts_contrib_fig}}
\end{figure}

\placefigure{counts_contrib_fig}

\section{Discussion on evolution}
\label{evolution_sec}

As discussed in Section \ref{result_sec}, the redshift range sampled by the
galaxies used to determine the LF of the two populations is
too narrow (0.0 $< z \le$ 0.4) to constrain the evolution rates with a high degree of
confidence. This is shown in Figure \ref{counts_contrib_fig}, where the contribution of the galaxies used in the
fitting procedure (dashed line) and of the unidentified
objects (dot-dashed line) to the total observed source counts (solid line) are
illustrated separately. The spectroscopic sample contributes only
marginally to the large excess with respect to the Euclidean expectations at $\sim$ 1mJy, which
is indeed dominated by optically unidentified sources. For this reason, to
test our evolution rates, we have compared the predictions of our model with
other observables, extending the analysis to lower flux densities and higher
$z$ than those reached by our survey (using the source counts and the HDF-N 
data-set). We have extrapolated the model results to higher redshift, given the evidences 
reported in Section \ref{unid_sec} that all the 15-$\mu$m sources
belong to the same population of galaxies.

\subsection{Comparison between model predictions and data}
 
We find that the agreement between the model predictions and the observables
is very good. In particular, the starburst population must evolve with the
evolution rates found ($k_l \sim 3.5$ and $k_d \sim 3.8$, see Table
\ref{fdl_tab}) up to $z_{break} \sim 1$ and no additional evolution
at $z > z_{break}$. Of course, the $k_l$ and $k_d$ values depend on how the different galaxy populations emitting in
the Mid-IR band are separated. However, once the subdivision is fixed, the evolution rates
are well determined, given the large number of observables to be fitted. 

\begin{figure}
\epsscale{0.5}
\plotone{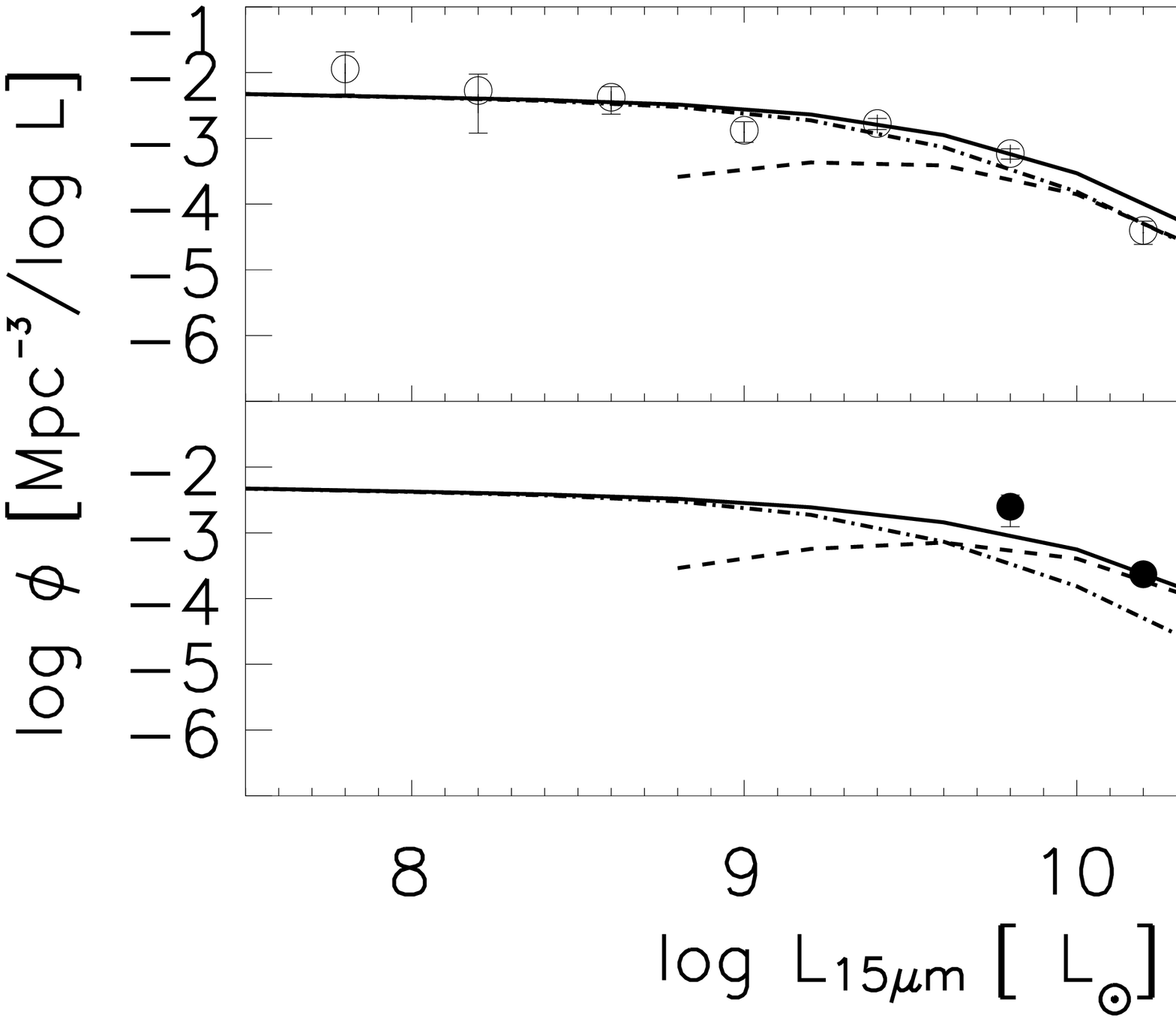}
\plotone{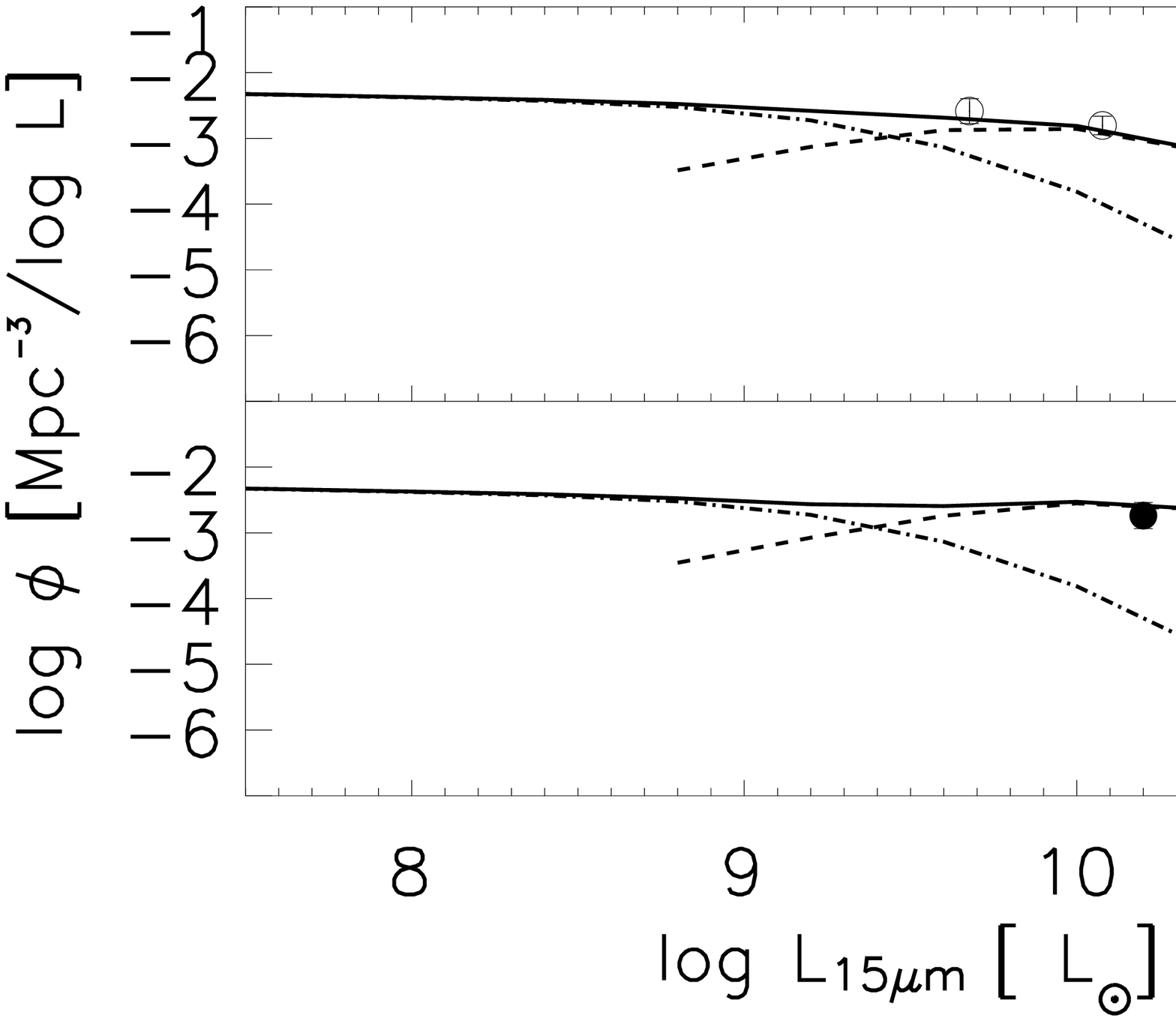}
\caption{{\it Top}: The rest-frame 15-$\mu$m
luminosity functions for the total sample of galaxies in the southern ELAIS
fields (S1+S2). Dashed, dashed-dotted and solid lines represent the starburst,
normal galaxies and total LFs from our model. {\it upper plot}: $0.0<z \le 0.2$
($z_{\rm mean}{\simeq}0.12$). {\it lower plot}: $0.2<z \le0.4$ ($z_{\rm
mean}{\simeq}0.27$). {\it Bottom}: The predicted rest-frame 15-$\mu$m
luminosity functions extrapolated to higher redshift intervals and compared
with data from the HDF-N survey. The points have been derived by
\cite{2003ApJ...587...90X} from $1/V_{max}$ analysis (private communication). {\it upper plot}: $0.4<z<0.7$ ($z_{\rm
  mean}{\simeq}0.55$). {\it lower plot}: $0.7<z<1.0$ ($z_{\rm
  mean}{\simeq}0.85$). \label{prev_fig}}
\end{figure}

In Figures \ref{prev_fig}a,b the total (normal spiral+starburst) luminosity functions predicted by our model in
the ELAIS southern fields (our sample) and in the HDF-N field, are compared with
the data. For the shallower ELAIS fields, the observed and predicted
luminosity functions have been computed in two low-redshift bins: $0.0<z\le0.2$ ($z_{\rm
mean}{\sim}0.12$) and $0.2<z \le0.4$ ($z_{\rm mean}{\sim}0.27$). For the deeper HDF-N, 
the observed and predicted luminosity functions are computed in two higher-redshift bins: $0.4<z<0.7$ ($z_{\rm
mean}{\sim}0.55$) and $0.7<z<1.0$ ($z_{\rm mean}{\sim}0.85$). In the latter
redshift bins, the model is an extrapolation to higher redshift
of our best-fitting model; the data points are from
\cite{2003ApJ...587...90X} (private communication), where
they were computed using the $V_{max}$ formalism. 

\begin{figure}
\epsscale{0.55}
\plotone{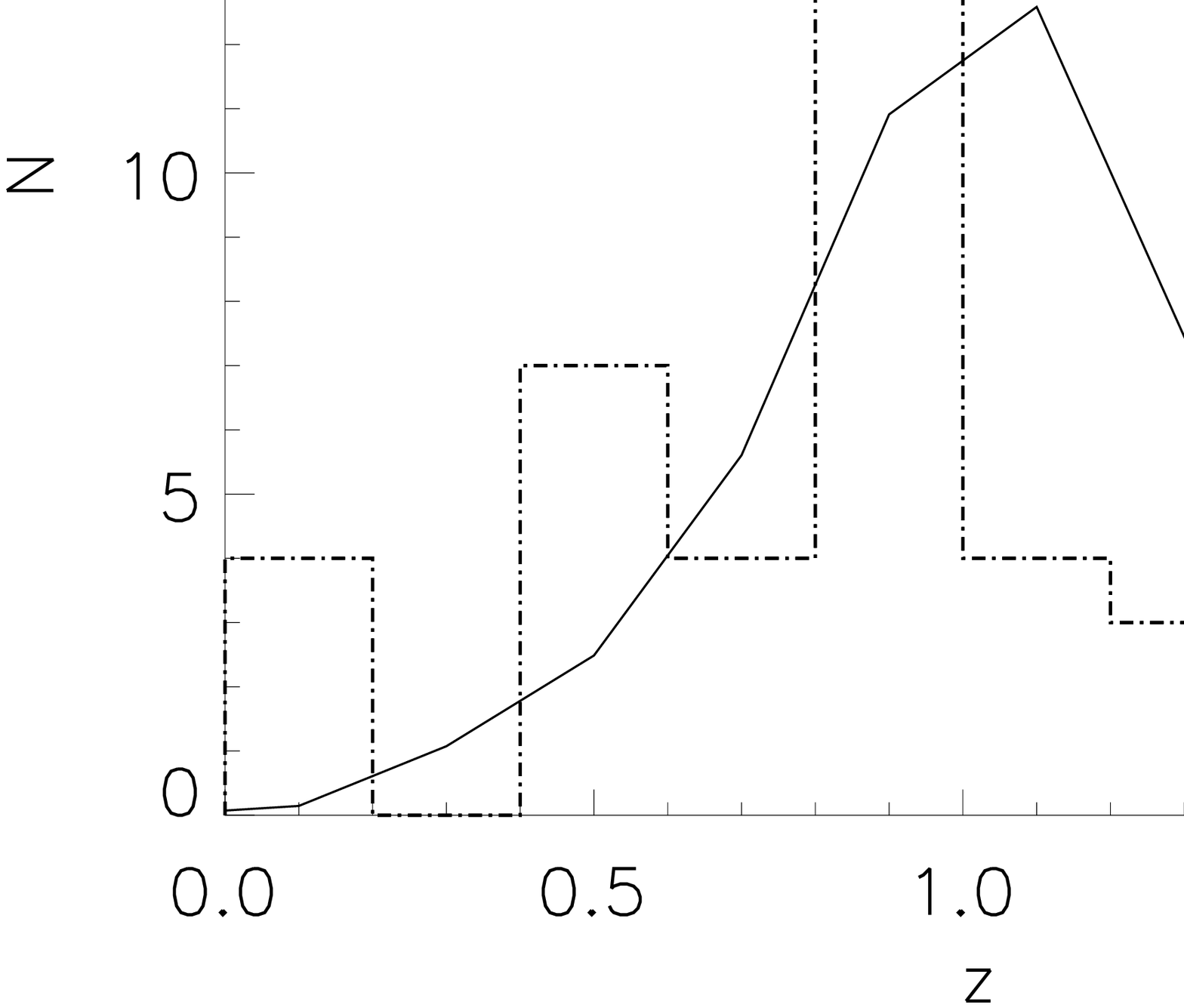}
\caption{Comparison between the observed
(dot-dashed histogram) and the predicted redshift distributions
(solid-line) in the HDF-N field. Sources with $S{\ge}0.10$ mJy have
been considered (41 objects; H. Aussel and S. Berta, 2003, private communication). \label{prev2_fig}}
\end{figure}

In Figure \ref{prev2_fig} the redshift distribution of sources with
$S>0.1$ mJy predicted by our model is superimposed to the data to the same flux
density limit observed in the HDF-N. The agreement between the observed and
the modeled distributions is very good, both in shape and normalization, with
both distributions showing a peak around $z\sim 0.9-1.0$.

\placefigure{prev_fig}

\placefigure{prev2_fig}

\begin{figure}
\epsscale{0.6}
\plotone{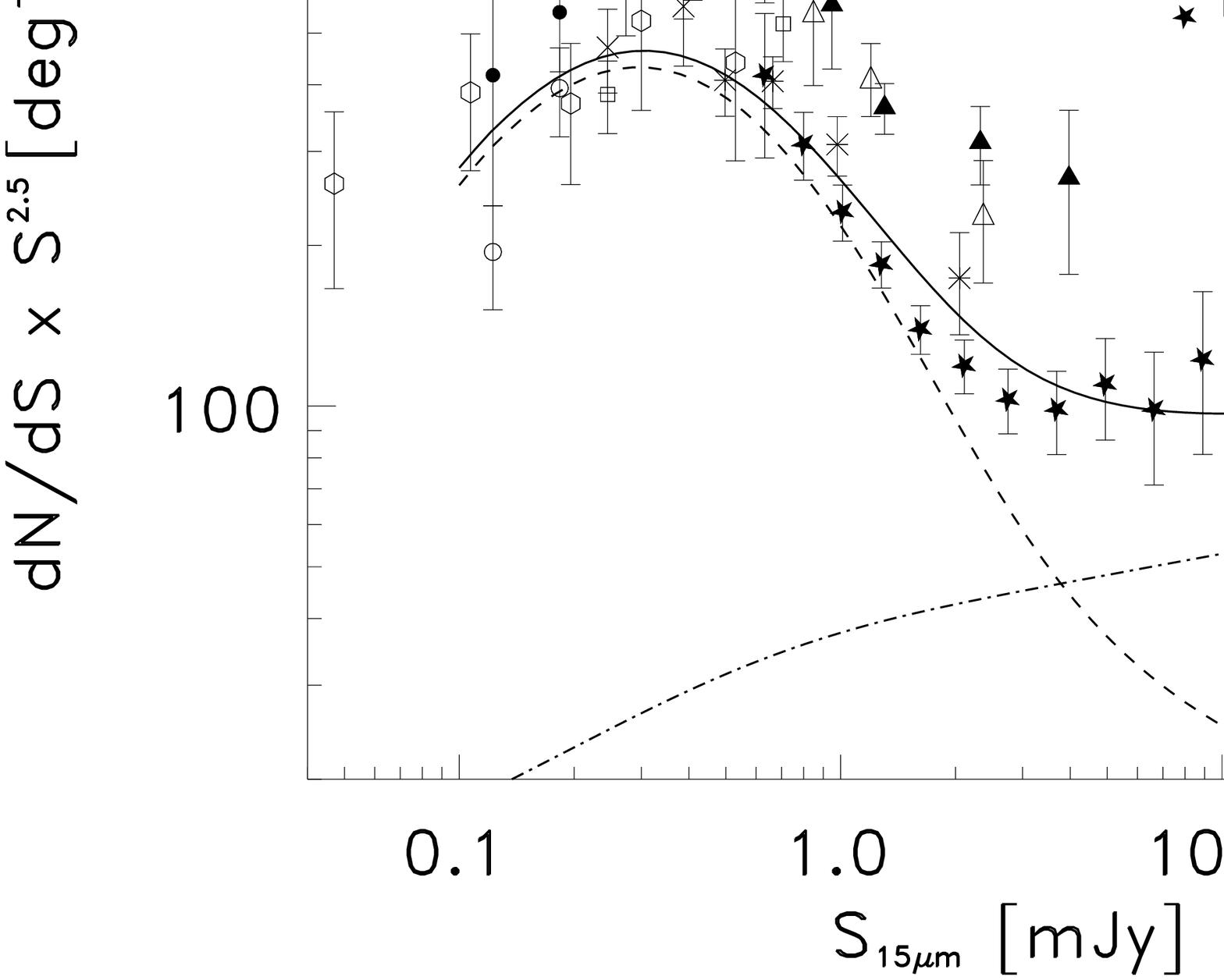}
\caption{Comparison of model prediction and observed 15-$\mu$m source
counts. Data points: the ELAIS-S1 counts reported by
\cite{2002MNRAS.335..831G} are plotted as filled stars. The new A2390 source
counts from \cite{2003A&A...407..791M} are plotted as open diamonds.
 Other data points from \cite{1999A&A...351L..37E}: ISO HDF-N (open circles), ISO HDF-S (filled circles), Marano Firback (open squares),
Marano Ultra-Deep (diagonal crosses), Marano Deep (asterisk), Lockman
Deep (open triangles), Lockman Shallow (filled triangles). Dashed and
dashed-dotted lines represent the contributions from starburst and normal
galaxies computed with our model. \label{counts_fig}}
\end{figure}

\placefigure{counts_fig}

In Figure \ref{counts_fig} the observed and the predicted differential source
counts are compared. Our model well reproduces the trend observed from the
IRAS flux densities down to
the ultra-deep survey limits. It is in perfect agreement with our ELAIS data
at fluxes $S_{15{\mu}m}{\lsimeq}1$ mJy and ${\gsimeq}3$ mJy, 
while it is slightly higher and smoother than our data in the
critical interval $1{\lsimeq}S_{15{\mu}m}{\lsimeq}3$ mJy, where
the counts start diverging from no evolution expectations and data from
different surveys show the larger differences. Our model is however 
intermediate between, and consistent within the errors with both ELAIS data and those of the other survey as reported by \cite{1999A&A...351L..37E} and \cite{2003A&A...407..791M}.
The sharp upturn shown by the ELAIS source counts around 2 mJy
could be well reproduced by \cite{2002MNRAS.335..831G} by introducing
a luminosity cut-off in the local luminosity function of starburst galaxies at $L=10^{10.8} 
L_{\odot}$. Although our data cannot either confirm or rule out
this hypothesis, in the present work we have chosen not to 
introduce any ``artificial'' constraint in our maximum likelihood
luminosity function determination.

\subsection{Comparison with other evolutionary models}

The idea of modeling the Mid-IR source counts and luminosity function by
dividing the sources into different populations, following
different evolutionary schemes, was first proposed by 
\cite{2001A&A...378....1F}. The \cite{2001A&A...378....1F} and our local
population subdivisions are similar (see Section \ref{llf_section}), although the evolutionary rates required by 
\cite{2001A&A...378....1F} for the starburst population
are slightly higher than ours: ${\sim}(1 + z)^{3.8}$ in luminosity and ${\sim}(1 + z)^4$ in density. For this reason, the source counts predicted by \cite{2001A&A...378....1F} are somewhat 
higher and smoother than ours, especially
in the flux density interval $1<S<5$ mJy, and their modeled redshift distribution, although similar at faint
flux densities (i.e. $S_{15{\mu}m}{\geq}0.1$ mJy in the HDF-N),
shows a significant peak around $z{\simeq}0.8-1$ even
at relatively high flux densities (${\gsimeq}2$ mJy and up
to 10 mJy), not observed in the data. Despite these differences, the
\cite{2001A&A...378....1F} model is the model which, among all those
existing in literature, produces results more similar to ours.

For example, the recent model of \cite{2003ApJ...587...90X}
predicts source counts that are not only significantly
higher than our expectations, but also higher than all the
15-$\mu$m source counts derived by the different Mid-IR
surveys, over a large flux density interval
($1{\lsimeq}S_{15{\mu}m}{\lsimeq}10$ mJy). 
The large discrepancy between the \cite{2003ApJ...587...90X} and our models
might be due to a combination of several causes, including a different
population subdivision of the local luminosity function (see Fig. \ref{llf_pop_fig}) and
different AGN contribution. 
As shown in the previous section, the local luminosity function of starburst
galaxies, derived by \cite{2003ApJ...587...90X} (see Fig. \ref{llf_pop_fig}) has a very pronounced high
luminosity tail in contrast with our determination. This tail causes the
high predicted counts at few mJys, since local galaxies,
 characterized by $L_{15{\mu}m}{\sim}10^{11}L_{\odot}$ and undergoing a
 moderate-high evolution, at a typical redshift of $z{\sim}1$ would have 
$L_{15{\mu}m}(z{\sim}1){\sim}8-10L_{15{\mu}m}(z=0)$ and flux densities
 in the mJy range. On the contrary, local starburst galaxies with typical
 luminosities of $L_{15{\mu}m}{\lsimeq}10^{9-10}L_{\odot}$ (as in our model)
 would have $L_{15{\mu}m}{\sim}10^{11}L_{\odot}$ at $z \sim 1$ and expected fluxes in the sub-mJy
 range ($S\sim$0.4 mJy), where the bump observed in the
 differential counts is located. Moreover, the evolution considered by
 \cite{2003ApJ...587...90X} for AGN is higher than that found by I. Matute (private communication).

Although all the models give rise to different
results, it is interesting to note that they all agree in the determination of
the total local luminosity function (see Section \ref{llf_section}),
though with different population subdivisions and/or evolution
hypotheses.
 
\subsection{Star formation History predicted by model}
 
The evolving luminosity function model can be used to determine the
star formation rate density of the Universe. We first calculate the luminosity
density at 15-$\mu$m by integrating the luminosity function over all the
luminosities; then we convert the luminosity
density into star-formation density by using the star-formation calibrator
based on Mid-IR data. To this purpose, we have used the calibration given by
\cite{2002MNRAS.332..549M} for a \cite{1955VA......1..283S} IMF, over the mass range
[0.1,100]M$_{\odot}$. Since the \cite{2002MNRAS.332..549M} infrared estimator is based on the $L_{60{\mu}m}$ bolometric luminosity
 (SFR(M$_{\odot}$yr$^{-1}$)=${\lambda}L_{\lambda}(60{\mu}m)/1.5{\cdot}10^{36}$(W)), 
 $L_{15{\mu}m}$ has been converted to $L_{60{\mu}m}$ following
 \cite{2001NewA....6..265M} ($L_{60{\mu}m}/L_{15{\mu}m}\sim$5). The
 adopted value is consistent within $\sim$20-30 $\%$ with the values
 of M51 and M82 (see Fig. \ref{sed_fig}). In the same way, we have also translated into SFR density the $1/V_{max}$ luminosity
function results derived in two redshift bins
(0.0 $< z \le$ 0.2 and 0.2 $< z \le$ 0.4).

\begin{figure}
\epsscale{0.55}
\plotone{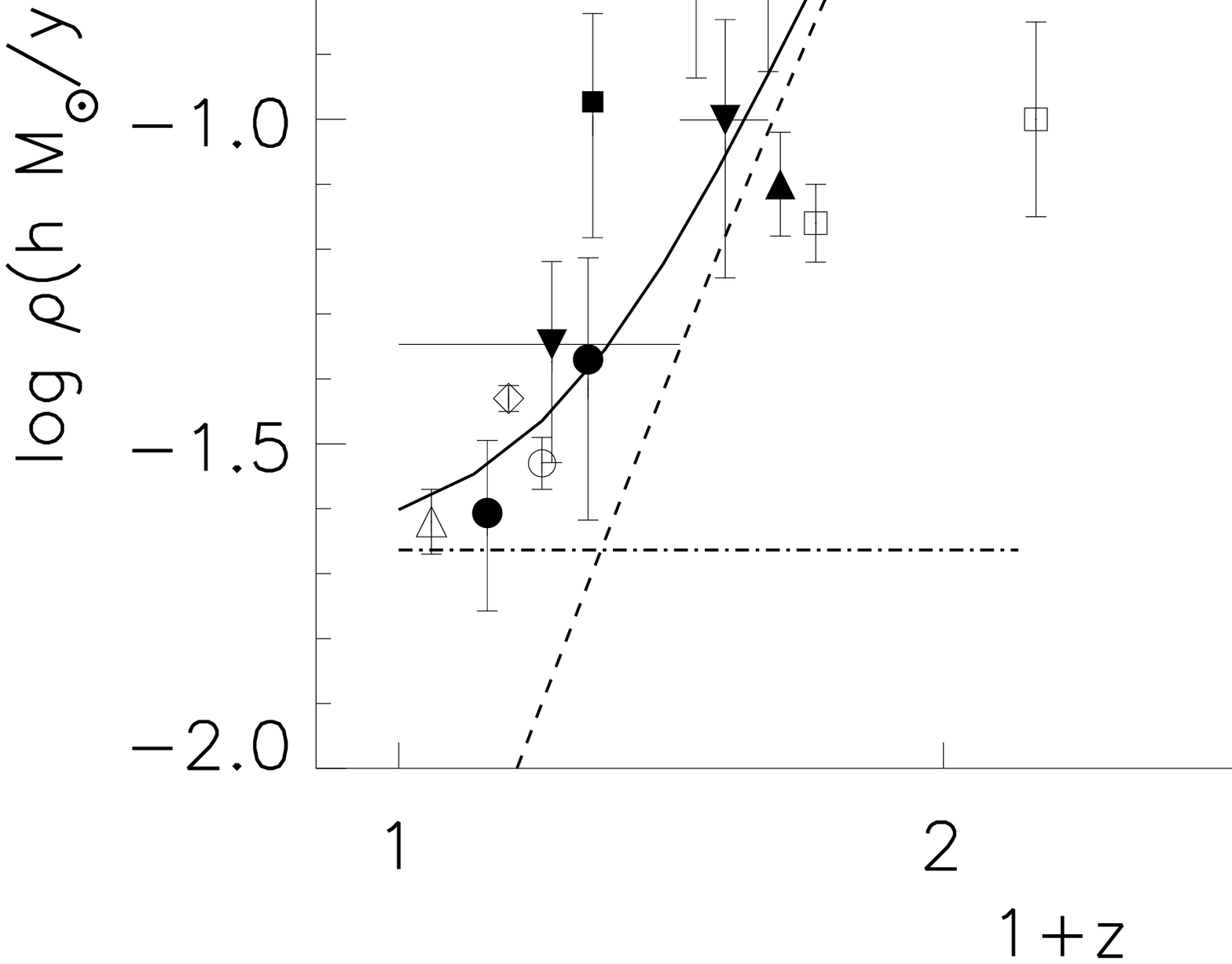}
\caption{A compilation of constraints on the star formation history of the
  Universe mainly taken from \cite{2001MNRAS.320..504S}. The UV data have been corrected
  for dust extinction following \cite{2001MNRAS.320..504S}, while the
  H$\alpha$ data have been corrected by the original authors. In all cases a $\Omega_m =
  0.3$, $\Omega_{\Lambda} = 0.7$ cosmology and a
  \cite{1955VA......1..283S} IMF, over
  the mass range [0.1,100]M$_\odot$, were assumed. Units are hM$_\odot$yr$^{-1}$Mpc$^{-3}$. 
The filled circles and the solid curve are from this work (1/V$_{max}$ analysis
of ELAIS data and prediction of our maximum likelihood model, the dashed and the dot-dashed lines being the
contribution of starbursts and spirals respectively).  The symbols are as
  follows: empty upward-pointing triangles - \cite{1998yugf.conf...96G}; empty
  circles - \cite{1998ApJ...495..691T}; empty diamonds - \cite{1998MNRAS.300..303T};
 filled downward-pointing triangles - \cite{2002MNRAS.332..549M};
 filled upward-pointing triangles - \cite{1999ApJ...517..148F}; filled squares - \cite{2000ApJ...544..641H};
  empty squares - \cite{1996AJ....112..839C}; empty downward-pointing
  triangles - \cite{1999ApJ...519....1S}; filled stars - \cite{1998Natur.394..241H}. \label{sfr_fig}}
\end{figure}

In Figure \ref{sfr_fig} we show a compilation of estimates on the star
formation rate density as a function of redshift from different indicators taken from
\cite{2001MNRAS.320..504S}, in a $\Omega_m = 0.3$, $\Omega_{\Lambda} = 0.7$
cosmology in units of hM$_\odot$yr$^{-1}$Mpc$^{-3}$. The UV data have
been corrected for dust extinction following \cite{2001MNRAS.320..504S}, while
the H$\alpha$ data have been corrected for dust extinction by the original authors. The
estimates derived from Mid-IR data in the HDF-S by \cite{2002MNRAS.332..549M} and from radio data by
\cite{2000ApJ...544..641H} have also been added, after conversion to the
adopted cosmology. 
The prediction of our model is shown as a solid line and the $1/V_{max}$ data points as filled circles.

Our model predicts a trend for the star-formation density similar to the
results obtained in other bands, with a rapid increase from $z\sim$0 to $z\sim$1, followed
by a flat plateau at high $z$. The actual measured data points at $z \le 0.4$ are of particular interest, since they provide an estimate of the
star-formation density at a relatively low redshift, but not so local to be affected by
clustering and local dishomogeneities. We found star formation density of
$\dot \rho$=(0.025$\pm$0.007)hM$_\odot$yr$^{-1}$Mpc$^{-3}$ and $\dot \rho$=(0.043$\pm$0.020)hM$_\odot$yr$^{-1}$Mpc$^{-3}$ at the two average redshifts $z=0.12$ and
$z=0.27$, respectively. 
At $0.4{\lsimeq}z{\lsimeq}$1.0, where the ELAIS data are highly incomplete, the model 
has been constrained by other high-$z$ observables in literature.
We do not extrapolate the predictions of our model at $z >$1.3, since the LW3
ISO filter does not allow to efficiently sample this range of redshift (see LW3 K-correction in the Fig. \ref{sed_fig}).
Our model prediction is consistent with the estimates derived from
UV, optical and Mid-IR data up to $z \le 0.4$. In particular, our results are in excellent
agreement with the data result obtained by \cite{2002MNRAS.332..549M} from the Mid-IR survey in the HDF-S
(filled downward-pointing triangles). 
At $0.4 \lsimeq z \lsimeq 1.0$, the model is significantly higher than the
extinction corrected UV data, suggesting that the extinction corrections
applied could be underestimated at those redshifts. The estimates derived from radio
data by \cite{2000ApJ...544..641H} (filled squares) are sistematically
higher than our model predictions by about a factor of two. However, these data might be overestimated, since the authors considered
as star-forming galaxies (thus contributing to the star-formation density) also all the unidentified 
radio sources, even at flux densities where the fraction of elliptical radio
galaxies could still be significant (see \citealp{1999MNRAS.304..199G}).

\placefigure{sfr_fig}

\section{Conclusion}
\label{concl_sec}

We have presented the first direct determination of the 15-$\mu$m luminosity
function and its cosmic evolution for galaxies. 

As previously found by other authors, three populations of sources give rise to
the 15-$\mu$m emission: the normal, the starburst and the AGN
populations, characterized by different cosmic evolution. In this work we have
analysed the galaxy component only (quiescent plus actively starforming). 
The contribution of the AGN component is discussed in a separate paper
(\citealp{2002MNRAS.332L..11M}; Matute et
al. in preparation). 

The analysis is based on data from the
ELAIS southern fields survey. The sample is composed by $\sim$150 galaxies
 in the redshift interval 0.0$<$z$\le$0.4 and covers a large flux density range intermediate
between the IRAS and the deep ISOCAM surveys (0.5$\le{S}\le{50}$ mJy). 
Differently from other authors, we have adopted in this work the
$L_{15{\mu}m}/L_{\rm R}$ ratio as a criterion to separate the quiescent, non-evolving and the starburst, evolving populations.
 This criterion, suggested by the existing correlation between
 $L_{15{\mu}m}/L_{\rm R}$ and the amount of activity in galaxies, is a
 posteriori supported by the results of the V/V$_{max}$ analysis on the two
 populations defined on this basis.

The main results of our analysis are:

\begin{enumerate}

\item{In the ML analysis we have simultaneously fitted
  both the evolution rates and the shape parameter of the local
  LF for both the spiral and the starburst populations. We have assumed that the spiral population does not evolve, while we have
  let to evolve the starburst population both in luminosity and in
  density. Since the two populations sample different
  luminosity ranges, we have obtained an accurate determination for the faint end
  of the LLF for the
  quiescent component, while the knee and the $\sigma$ parameters of the LLF
  are better constrained for the starburst one. The evolution found for the active population is
  $\sim(1+z)^{3.5}$ in luminosity and $\sim(1+z)^{3.8}$ in density, up to $z_{break}\sim1$.}

\item{Our total 15-$\mu$m LLF is in agreement with previous
    determinations derived from the IRAS data. On the contrary, the LLFs for different populations derived by different
  authors have not the same level of consistency. While in our subdivision the
  quiescent population is expected to dominate locally over all the
  luminosity range, in other models (i.e. \citealp{2003ApJ...587...90X}) the starburst population
  dominates locally at high luminosities, leading to a large discrepancy in
  the model predictions.}

\item{To test the evolution parameters with higher degree of confidence, we
    have compared our model predictions with all the observables
 existing in literature, over all the $z$ and flux ranges (source counts,
 luminosity functions, $z$-distributions). Our best-fitting model well
 reproduces all the observables. In the critical interval 1${\lsimeq}S{\lsimeq}$3 mJy,
 where the source counts from different surveys show the larger discrepancies, our
 model is intermediate between the data from ELAIS-S1 \citep{2002MNRAS.335..831G}
 and the data from the deep surveys \citep{1999A&A...351L..37E}. On the other hand,
 in the flux range 3${\lsimeq}S{\lsimeq}$10 mJy, our
 model is well consistent with existing data, differently from the
 \cite{2003ApJ...587...90X} model, whose predicted differential sources counts
 are at least a factor of 3 higher than the data.}

\item{Using the evolutionary model found for the 15-$\mu$m galaxies and the data
    points from the $1/V_{max}$ LF analysis, we have
  estimated the star-formation rate density. The redshift range sampled by our data
  ($0.0 < z \le0.4$) is of particular interest, since it provides an estimate of
  the star-formation at relatively low redshift, but not so local to be affected by
clustering and local dishomogeneities. We find $\dot \rho$=(0.025$\pm$0.007)
hM$_\odot$yr$^{-1}$Mpc$^{-3}$ and $\dot \rho$=(0.043$\pm$0.020)
hM$_\odot$yr$^{-1}$Mpc$^{-3}$ at the two mean redshifts $z=0.12$ and
$z=0.27$, respectively. At $z \lsimeq 0.4$ our model predictions 
are well consistent with other estimates derived from UV, optical and
    Mid-IR data. At higher redshift our model predictions are
    significantly higher than the UV extinction corrected data and
    lower by about a factor of two than the estimates derived from radio data by
\cite{2000ApJ...544..641H}.}

\end{enumerate}

\acknowledgments

F.P. was partially supported by the European Program HUMAN POTENTIAL (Contract
Number HPMT-CP-2000-00096). F.P. would like to thank Lucia Pozzetti for helpful 
discussion.

\bibliographystyle{apj}
\bibliography{aamnemonic,bib}

\clearpage

\begin{table}
\begin{center} 
\caption{15-$\mu$m Luminosity Function parameters from the ML analysis.}
\label{fdl_tab}
\begin{tabular}{ccccccc} \\\hline 
Population & $\alpha$           &$\sigma$ & $\log{L_{\star}}$
&$\log\phi^{\star}$ &$k_l$ &$k_d$  \\\hline
normal spirals   &1.10$^{+0.25}_{-0.25}$&0.5$^{+0.1}_{-0.2}$&8.8$^{+0.7}_{-0.9}$ &
-2.45 & 0.0  & 0.0\\
           &                      &                   &                    &
& & \\
starbursts&0.0 (fixed) &0.39$^{+0.025}_{-0.025}$&8.8$^{+0.3}_{-0.2}$& -3.53 & 3.5$^{+1.0}_{-3.5}$ &3.8$^{+2.0}_{-2.0}$\\\hline
\end{tabular}
\end{center}
\end{table}

\clearpage

\begin{table}
\begin{center} 
\caption{15-$\mu$m Local luminosity Function from the 1/V$_{max}$ analysis.}
\label{fdl_vmax_tab}
\begin{tabular}{ccc} \\\hline 
 $\log[{\nu}L_{\nu}/L_{\odot}]$ & $\log[\phi(Mpc^{-3}mag^{-1}]$ & 1$\sigma$ Error \\ \hline
  7.8   &  -1.94  &  (+0.52)(-0.76)\\
  8.2   &  -2.27  &  (+0.24)(-0.64) \\
  8.6   &  -2.37  &  (+0.16)(-0.25)\\
  9.0   &  -2.84  &  (+0.12)(-0.17)\\
  9.4   &  -2.81  &  (+0.08)(-0.09)\\
  9.8   &  -3.24  &  (+0.06)(-0.08)\\
 10.2   &  -4.25  &  (+0.11)(-0.14)\\
 10.6   &  -5.21  &  (+0.23)(-0.53)\\\hline

\end{tabular}
\end{center}
\footnotesize
\end{table}

\clearpage

\clearpage 

\clearpage 

\clearpage 

\clearpage

\clearpage

\clearpage

\clearpage 
\vspace{3cm}

\clearpage 

\clearpage 

\clearpage 

\clearpage

\clearpage

\clearpage

\end{document}